\documentclass[5p,authoryear]{elsarticle}
\usepackage{graphicx}
\usepackage{subfigure}

\usepackage{pdfpages}
\usepackage{helvet}
\usepackage{courier}
\usepackage{url}
\usepackage{lscape}
\usepackage{booktabs}
\usepackage{multirow}
\usepackage{amsmath}
\usepackage{mathtools}
\usepackage[ruled, vlined, linesnumbered]{algorithm2e}
\usepackage{wrapfig}

\makeatletter
\def\BState{\State\hskip-\ALG@thistlm}
\makeatother

\usepackage{array}
\usepackage{ragged2e}

\newcolumntype{P}[1]{>{\centering\arraybackslash}p{#1}}
\journal{Expert Systems with Applications}

\makeatletter
\def\@author#1{\g@addto@macro\elsauthors{\normalsize%
    \def\baselinestretch{1}%
    \upshape\authorsep#1\unskip\textsuperscript{%
      \ifx\@fnmark\@empty\else\unskip\sep\@fnmark\let\sep=,\fi
      \ifx\@corref\@empty\else\unskip\sep\@corref\let\sep=,\fi
      }%
    \def\authorsep{\unskip,\space}%
    \global\let\@fnmark\@empty
    \global\let\@corref\@empty  
    \global\let\sep\@empty}%
    \@eadauthor={#1}
}
\makeatother

\begin{document}


\begin{frontmatter}
\title{Investigating the Effectiveness of Representations Based on Pretrained Transformer-based Language Models in Active Learning for Labelling Text Datasets}
\author{Jinghui Lu}
\ead{Jinghui.lu@ucd.connect.ie}


\author{Brian Mac Namee}
\ead{Brian.MacNamee@ucd.ie}

\address{School of Computer Science, University College Dublin, Ireland}
\cortext[cor1]{Corresponding author}

\begin{abstract}
Manually labelling large collections of text data is a time-consuming and expensive task, but one that is necessary to support machine learning based on text datasets. \emph{Active learning} has been shown to be an effective way to alleviate some of the effort required in utilising large collections of unlabelled data for machine learning tasks without needing to fully label them. The representation mechanism used to represent text documents when performing active learning, however, has a significant influence on how effective the process will be. While simple vector representations such as \emph{bag-of-words} and embedding-based representations based on techniques such as \emph{word2vec} have been shown to be an effective way to represent documents during active learning, the emergence of representation mechanisms based on the pre-trained transformer-based neural network models popular in natural language processing research (e.g. \emph{BERT, GPT-2, XLNet}) offer a promising, and as yet not fully explored, alternative. This paper describes a comprehensive evaluation of the effectiveness of representations based on pre-trained transformer-based language models for active learning. This evaluation shows that  transformer-based models, especially BERT-like models, that have not yet been widely used in active learning, achieve a significant improvement over more commonly used vector representations like bag-of-words or other classical word embeddings like word2vec. This paper also investigates the effectiveness of representations based on variants of BERT such as Roberta, DistilBert, and Albert as well as comparing the effectiveness of  the ``[CLS]'' token representation and the aggregated representation that can be generated using BERT-like models. Finally, we propose an approach to tune the representations generated by BERT-like transformer models during the active learning process, \emph{Adaptive Tuning Active Learning}. Our experiments show that the limited label information acquired in active learning can not only be used for training a classifier but can also adaptively improve the embeddings generated by the BERT-like language models as well.
\end{abstract}

\begin{keyword}
active learning \sep word embeddings \sep Transformer \sep BERT \sep text labelling
\end{keyword}
\end{frontmatter}
\section{Introduction}\label{sec:intro}

\sloppy

 \emph{Active learning} (AL) \citep{settles2009active} is a semi-supervised machine learning technique that minimises the amount of labelled data required to build accurate prediction models. In active learning only the most informative instances from an unlabelled dataset are selected to be labelled by an oracle (i.e. a human annotator) to expedite the learning procedure. This property makes active learning attractive in scenarios where unlabelled data may be abundant but labelled data is expensive to obtain---for example image classification \citep{tong2001support,zhang2002active}, speech recognition \citep{tur2005combining}, and text classification \citep{hoi2006large,Liere1997active,zhang2017active,singh2018improving}. The use of active learning for text classification is the focus of this work. One crucial component in active learning systems for text classification is the mechanism used to represent documents in the tabular structure required by most machine learning algorithms. 

Vectorized representations based on word frequencies, such as \emph{bag-of-words} (BOW), are the most commonly used representations  in active learning \citep{singh2018improving,hu2010egal,hu2008sweetening,wallace2010active,siddhant2018deep,miwa2014reducing}. Considerable recent work, however, has shown that representations of natural language based on learned \emph{word embeddings} can be useful for a wide range of natural language processing (NLP) tasks including text classification \citep{mikolov2013efficient,pennington2014glove,bojanowski2017enriching,howard2018universal,radford2018improving,devlin2018bert,peters2018deep}. Standard approaches to learning word embeddings like \emph{word2vec} \citep{mikolov2013efficient}, \emph{Glove} \citep{pennington2014glove}, and \emph{FastText} \citep{bojanowski2017enriching,joulin2016bag}, or contextualized approaches such as \emph{Cove} \citep{mccann2017learned} and \emph{ElMo} \citep{peters2018deep} convert words to fixed-length dense vectors that capture semantic and syntactic features, and allow more complex structures (like sentences, paragraphs and documents) to be encoded as aggregates of these vectors. 

More recently document-level approaches have been developed, such as \emph{ULM-Fit} \citep{howard2018universal}, \emph{OpenAI GPT} \citep{radford2018improving}, and \emph{BERT} \citep{devlin2018bert}, that are pre-trained with large-scale generic corpora and then fine tuned to a specific task. The use of these approaches has been shown to significantly increase the performance in many downstream NLP tasks \citep{devlin2018bert,liu2019roberta,radford2019language}, and has also been shown to be useful for transferring knowledge learned from large generic corpora to downstream tasks focused on much more specific corpora. Transformer-based pre-trained models such as the \emph{bidirectional encoder representations from transformers}  (BERT) model have achieved particularly  impressive results across many NLP tasks \citep{devlin2018bert}. Even though word embeddings and  transformer-based language models have been widely applied in text classification, there is little work devoted to leveraging them in active learning for text classification \citep{zhang2017active,zhao2017deep,siddhant2018deep}, and a comprehensive benchmark comparison of their usefulness for active learning does not exist in the literature.

To address this gap in the literature this paper addresses four research questions:

\begin{description}

\item{\textbf{RQ1:}} \textit{Are representations generated using  pre-trained transformer-based language models more effective than other more commonly used representations in the context of active learning for text labelling? If so, which pre-trained model generates the most effective representations?}

\item{\textbf{RQ2:}} \textit{Can lightweight versions of transformer-based models be used instead of the standard large models to reduce the computational burden during active learning while still maintaining high performance levels?}

\item{\textbf{RQ3:}} \textit{When using embeddings generated using a transformer-based model is it more effective to represent a document using an aggregate of the word level embeddings produced by the model or to use the embedding of the ``[CLS]'' token?}

\item{\textbf{RQ4:}} \textit{Can we further improve the performance of an active learning system using a transformer-based model by \emph{fine tuning} the model?}

\end{description}


To answer these research questions, this paper describes a comprehensive evaluation experiment that explores the effectiveness of various text representation techniques for active learning in a text classification context. This evaluation, based on 8 datasets from different domains, including product reviews, news articles, and blog posts, shows that representations based on pre-trained transformer-based language models---and especially representations based on Roberta---consistently outperform the more commonly used vector representations such as word embeddings or bag-of-words. This demonstrates the effectiveness of transformer-based representations for active learning. 

Based on this first results we also report further studies using the same datasets that compare the performance of full and lightweight versions of the transformer-based models examined; that compare the performance of systems using the ``[CLS]'' token document representation and aggregated words document representation; and then we show that by fine tuning transformer-based models during active learning the performance of the active learning process can be improved. Taken together these experiments illustrate that by using text representations from transformer-based models some of the promise of deep learning \citep{lecun2015deep} can be brought to active learning while avoiding the considerable practical challenges of placing a deep neural network at the heart of the active learning process \citep{zhang2017active,zhao2017deep,siddhant2018deep,zhang2019neural}.  
7

The remainder of the paper is organized as follows: Section \ref{sec:rel_work} describes pool-based active learning, the text representation techniques used in this paper, and existing work related to the use of transformer-based models in active learning; Section \ref{sec:experiments_1} describes the experiment comparing the effectiveness of different text representation techniques for active learning; Section \ref{sec:experiment_2} describes the experiment that compares the performance of the active learning process when lightweight and complete variants of BERT are used  as well as the experiment comparing the use of the ``[CLS]'' token representation and aggregate representations; Section \ref{sec:experiment_3} shows the impact  that fine tuning the transformer-based models has on the active learning process; and, finally, Section \ref{sec:conclusions} draws conclusions and suggests directions for future work.

\section{Related Work}\label{sec:rel_work}

In this section we explain what is meant by pool-based active learning, describe the different text representations that are used in the experiments described in this paper, and describe existing work that investigates the impact of using different ext representations in active learning.

\subsection{Pool-based Active Learning}

The goal of active learning is to utilise a large collection of unlabelled data for supervised learning with minimal human labelling effort. In \emph{pool-based} active learning, a small set of labelled instances is used to seed an initial labelled dataset, $\mathcal{L}$. Then, according to a particular \emph{selection strategy}, a batch of data to be presented to an oracle for labelling is chosen from the unlabelled data pool, $\mathcal{U}$. After labelling, these newly labelled instances will be removed from $\mathcal{U}$ and appended to $\mathcal{L}$. This process repeats until a predefined stopping criterion has been met  (for example a label budget has been exhausted). 
 
The instances labelled by the oracle through this process can be used to train a predictive model. This model might be the final output of  the overall process, or this model can be used to generate labels for the remaining unlabelled instances in $\mathcal{U}$ to generate a fully labelled dataset with minimal effort from  the oracle. We mainly consider this latter scenario in this paper.

The selection strategy used to pick the unlabelled instances that will be presented to the oracle for labelling plays a vital role in active learning. Many different selection strategies are described in the literature. \emph{Model-based} selection strategies---such as uncertainty sampling \citep{lewis1994sequential} and query-by-committee \citep{seung1992query}---utilise models trained with the currently labelled instances, $\mathcal{L}$, to infer the ``\emph{informativeness}'' of unlabelled instances from $\mathcal{U}$. A small batch  of the most informative instances from the unlabelled pool are presented to the oracle for labelling at each iteration of the active learning process. \emph{Model-free} selection strategies---such as exploration guided active learning (EGAL) \citep{hu2010egal}---rely entirely on the features of instances in $\mathcal{L}$ and $\mathcal{U}$ to compute the ``\emph{informativenes}'' of each unlabelled instance without requiring the construction of a predictive model. 

Although a comparison of the effectiveness  of different selection strategies is not the focus of this paper, we use several commonly used selection strategies in our experiments to mitigate the impact of selection strategies when comparing the effectiveness of text representations. In the experiments described in this paper we use \emph{random sampling} (sample i.i.d from $\mathcal{U}$), \emph{uncertainty sampling} \citep{lewis1994sequential}, \emph{query-by-committee} \citep{seung1992query}, \emph{information-density (ID)} \citep{settles2008analysis}, and \emph{EGAL} \citep{hu2010egal} to alleviate the influence caused by different selection strategies.

\subsection{Text Representations}

The technique used to represent documents has a large impact on the effectiveness of active learning for text data. Representations schemes range from simple frequency based vector representations, like bag-of-words, to more sophisticated approaches based on word embeddings. This section describes the text representation schemes used in the experiments described in this paper. We describe simple frequency-based vector representations, word embedding representaitons, and transformer-based representations.

\subsubsection{Frequency-based Vector Representations}

\noindent \textbf{Bag-of-words} (BOW) is the most basic representation scheme for documents, and has been widely used in many active learning applications \citep{singh2018improving,hu2010egal,hu2008sweetening,wallace2010active,siddhant2018deep,miwa2014reducing}. Each column of a BOW vector records the number of times a word occurs in a document (known as \emph{term-frequency} (TF)), and 0 if the word is absent. The frequency of words is often weighted by inverse document frequency to penalise terms commonly used in most documents. This is known as TF-IDF \citep{sparck1972statistical}.

\noindent \textbf{Latent Dirichlet Allocation} (LDA) \citep{blei2003latent} is a topic modelling  technique designed to infer the distribution of membership to a set of topics across a set of documents. The model generates a term-topic matrix that captures the association between words and topics and a document-topic matrix that captures the association between a documents and topics. Each row of the document-topic matrix is a topic-based representation of a document where the $i$th column determines the degree of association between the $i$th topic and the document. This type of topic representation of documents has been used in active learning for labelling the relevance of studies to  systematic literature reviews \citep{singh2018improving,hashimoto2016topic,mo2015supporting}.

\subsubsection{Word Embedding Representations}

A word embedding is a word representation learned via mapping words into a vector space where the distance between words in that space is related to the semantic and syntactic features of the words. Word embeddings are typically learned by building prediction models that perform context prediction, for example, predicting the most likely target word given a set of context words. 

The word embedding representations investigated in this study are: \textbf{word2vec} \citep{mikolov2013efficient,mikolov2013distributed}, which was the earliest word embedding technique to garner widespread attention and is very commonly used; \textbf{Glove} \citep{pennington2014glove} which learns embeddings by approximating the relationship between word vectors and the word co-occurrence probabilities matrix; and \textbf{FastText} \citep{bojanowski2017enriching} which is a trained neural language model that enriches the training of word embeddings with subword information which improves the ability to obtain word embeddings of out-of-bag words. 

In our experiments, as is common practice, we average the vectors of words appeared in the document as the document representation. 

\subsubsection{Transformer-based Representations}

The transformer model was proposed by Vaswani et al. (\citeyear{vaswani2017attention}) and has since received massive attention from the NLP research community. The key idea behind transformer-based prediction models is extracting general language knowledge via pre-training the model with large unlabelled generic corpora, then fine-tuning the pre-trained model to fit the specific downstream tasks. It is possible, however, to use the document representations that arise from pre-training transformer models without fine-tuning. We investigate the effectiveness of embeddings produced by the most well-known transform-based models in this study.

 \textbf{BERT} \citep{devlin2018bert} has achieved amazing results in many NLP tasks. This model is trained with the plain text through masked language modeling (MLM) and next sentence prediction (NSP) tasks to enable a bidirectional learning of contextualized word embeddings. Contextualized word embedding implies a word can have different embeddings according to its context which alleviates the problems caused by polysemy etc. BERT was reported to achieve state-of-the-art results across 11 NLP tasks when it was proposed. Though it has since been out-performed by other language models, BERT is still regarded as an important step in NLP and most transformer-based models are variants of BERT or rely upon ideas from BERT.

 \textbf{Roberta} \citep{liu2019roberta} is a an optimized version of BERT where key hyper-parameters of the model are more  carefully selected, the training process is modified (the NSP task is removed and the MLM task is slightly modified), and the model is pre-trained with larger datasets than  those used to train BERT. Roberta is still near the top of leaderboard in GLUE benchmark. \footnote{https://gluebenchmark.com/leaderboard}

 \textbf{DistilBert} \citep{sanh2019distilbert} is a lightweight version of BERT in which knowledge distillation is leveraged to transfer generalization capabilities from \emph{regular} BERT. Sanh et al. (\citeyear{sanh2019distilbert}) show that ``\textit{it is possible to reduce the size of a BERT model by 40\%, while retaining 97\% of its language understanding capabilities and being 60\% faster}''.

 \textbf{Albert} \citep{lan2019albert} is another lightweight version of BERT that has lower memory requirements and higher training speed. Embedding matrix factorization and cross-layer parameter sharing are applied to reduce the number of parameters in the model. During training the NSP task is replaced with a harder sentence-order prediction task to maintain the generalisation ability of the language model. 

 \textbf{XLNet} \citep{yang2019xlnet} is another transformer-based model which has drawn much attention. It replaces the MLM training task used by BERT with a permutation language model task and uses a larger corpus than BERT for training. XLNet has been shown to outperform BERT on most tasks \citep{yang2019xlnet}, especially for tasks involving longer text sequences.

 \textbf{GPT-2} \citep{radford2019language} is a scaled-up version of OpenAI GPT model \citep{radford2018improving} which is also transformer-based. Different from the high-profile BERT-like model, GPT-2 adopted a unidirectional transformer with more layers (48 layers with 1.5 billion parameters). As compared to BERT, GPT-2 is pre-trained only via predicting the next word given previous words with a very large high-quality web text corpus across many domains. In our experiments, due to the GPU memory limitation, we adopted a small architecture of GPT-2.\footnote{\label{huggingface}https://github.com/huggingface/transformers}

\subsection{Using Word Representation Produced by Neural Network in Active Learning}

Although applying word embedding representations and transformer-based representations in text classification has attracted considerable attention in the literature \citep{mikolov2013efficient,bojanowski2017enriching,howard2018universal,devlin2018bert}, the use of these representations in active learning is not well explored. There are some studies proposed deep active learning procedure where word2vec are combined with convolutional neural networks (CNN)\citep{zhang2017active} or recurrent neural networks and gated recurrent units to predict the classes of documents \citep{zhao2017deep}. 
Very recently, Zhang Ye (\citeyear{zhang2019neural}) proposed a selection strategy that utilized the discrepancy between a basic BERT model and a  BERT model which is continuously pre-trained via the MLM task using a local corpus. However, the experiments deccribes by Zhang Ye focused on  selection strategies, rather than the impacts of different text representation techniques. 
Siddhant and Lipton (\citeyear{siddhant2018deep}) compare the performance of Glove-embedding-based active learning frameworks, which are composed of different classifiers such as bi-LSTM model and CNN, across many NLP tasks. They find that Glove embeddings selected by Bayesian Active Learning by Disagreement plus Monte Carlo Dropout or Bayes-by-Backprop Dropout usually outperforms the shallow baseline. However, Siddhant and Lipton take Linear SVM combined with BOW representation rather than Glove embeddings as a shallow baseline which makes the conclusion limited.  
Additionally, these studies focus on comparing the impact of selection strategies when used with deep neural networks, instead of that of text representations.


Apart from leveraging deep learning classifier and word embedding, some studies combine word embedding with low complexity machine learning algorithms such as Support Vector Machine (SVM). 
Hashimoto et al. (\citeyear{hashimoto2016topic}) propose a method, paragraph vector-based topic detection (PV-TD), that combines doc2vec \citep{le2014distributed} (an extension of word2vec) with k-means clustering to perform simple topic modelling. For the active learning process documents, which are represented by their distance to the cluster centres that result from the application of k-means, are fed into the SVM model. In their experiments PV-TD is shown to perform well compared to representations based on an LDA, and word2vec. 
Interestingly, Singh et al. (\citeyear{singh2018improving}) extend the experiments in \cite{hashimoto2016topic} with more datasets in the health domain, demonstrating that directly using doc2vec or BOW, rather than PV-TD, can achieve better results which is contrary to that obtained by Hashimoto.

Despite the promising results, these studies of active learning only using classical word embeddings explore a limited number of selection strategies (i.e. certainty sampling and certainty information gain sampling) and just focus on highly imbalanced datasets from specialist medical domains. 

To the best of the authors' knowledge, there is no research investigating the effectiveness of transformer-based-representations in active learning for text classification. This research fills this gap by comparing the effectiveness of various text representations in active learning in a benchmarking experiment that uses a range of selection strategies and datasets from multiple domains. 

\section{Comparing the Effectiveness of Text Representations}\label{sec:experiments_1}



This section describes the design of an experiment performed to evaluate the effectiveness of different text representation mechanisms in active learning, addressing \textbf{RQ1} from Section \ref{sec:intro}. To mitigate the influence of different selection strategies on the performance of the active learning process we also include a number of different selection strategies in the experiment. The following subsections describe the experimental framework, the configuration of the models used, the performance measures used to judge the effectiveness of different approaches, the datasets used in the experiments, and the experimental results and analysis.

\subsection{Experimental Framework} \label{sec:active_setting}

We apply pool-based active learning using different text representation techniques and selection strategies over several well-balanced fully labelled datasets. All datasets are from binary classification problems. The use of fully labelled datasets allows us to simulate data labelling by a human oracle, and is common in active learning research \citep{zhang2017active,zhao2017deep,singh2018improving,hu2010egal,hashimoto2016topic}. At the outset, we provide all learners with the same 10 instances (i.e. 5 positive instances and 5 negative instances) sampled i.i.d. at random from a dataset to seed the active learning process. Subsequently, 10 unlabelled instances, whose ground truth labels will be revealed to each learner, are selected according to a certain selection strategy. These examples are moved from $\mathcal{U}$ to $\mathcal{L}$ (with their labels) and the classifiers are retrained. We assume it is unrealistic to collect more than 1,000 labels from an oracle, and so we stop the procedure when an annotation budget of 1,000 labels is used up. As the batch size for selection is 10, this means that an experiment is composed of 100 \emph{rounds} of the active learning process which uses up the label budget of 1,000 labels. After each round the retrained classifier is used to label all of the examples remaining in $\mathcal{U}$. Algorithm \ref{alg:AL} shows the details of the active learning procedure using uncertainty sampling as an example selection strategy. In our experiment, this process is repeated 10 times using different random seeds. The performance measures reported are averaged across these repetitions.

\begin{algorithm}[h!]
\DontPrintSemicolon
\KwIn{$T$, set of all corpus \\
    $P$, index of all ground truth positive documents \\
    $N$, index of all ground truth negative documents \\
    $LM$, pre-trained language model}
\KwOut{$S$, set of accuracy+ scores of each loop\\
$L$, set of documents pseudo labelled by the oracle \\
 $R$, set of documents labelled by the classifier\\}
  
\SetKwBlock{Init}{Initialization}{}{}
\Init{  
        // \hspace{0.15cm}Infer document representations\\
        $E\leftarrow Inference(T,LM)$;\\
        // \hspace{0.15cm}Random sampling 5 neg and 5 pos\\
        $L\leftarrow Random(E,P,5)\cup Random(E,N,5)$;\\
        $\neg L\leftarrow E\setminus L$;\\
        $R\leftarrow \emptyset$;  \\
        $loop\leftarrow 0$;\\
        $Params\leftarrow \emptyset$;\\
        $S\leftarrow \emptyset$;
}
\SetKwRepeat{Do}{do}{while}
\While{$\left |L  \right | \le 1000$}{

 $CL,Params\leftarrow Train(loop,L,P,N,Params)$; \\
 $X\leftarrow Query(CL,\neg L)$;\\
 $L,\neg L,R\leftarrow Assign(CL,L,X)$; \\
 //\hspace{0.15cm} Compute accuracy+ score\\
 $S\leftarrow S\cup Eval(L,R,P,N)$;\\
 $loop\leftarrow loop + 1$;
}
\SetKwBlock{Begin}{Function}{end function}
\Begin($\text{Train} {(} loop,L,P,N,Params {)}$)
{

      \uIf{$loop\mod10 == 0$}
      {
        //\hspace{0.15cm} Cross-validation for hyper-parameters\\
        $Params\leftarrow CV(L,P,N)$; \\
        //\hspace{0.15cm} Train linear SVM\\
        $CL\leftarrow SVM(L,Params)$;
      }
      \Else
      {
        $CL\leftarrow SVM(L,Params)$
      }

  \Return{$CL,Params$}
}
\Begin($\text{Query} {(} CL,\neg L {)}$)
{
      //\hspace{0.15cm} Uncertainty sampling 10 examples\\
      $X\leftarrow argsort(abs(CL.deci\_func(\neg L)))[:10]$;

  \Return{$X$}
}
\Begin($\text{Assign} {(} CL,L,X {)}$)
{

      $L\leftarrow L\cup X$;\\
      $\neg L\leftarrow E\setminus L$;\\
      $R\leftarrow CL.predict(\neg L)$;

  \Return{$L,\neg L,R$}
}

\caption{Pseudo-code for active leaning.}\label{alg:AL}

\end{algorithm}

The classifiers used in the active learning process in all of our experiments are Linear-SVM models\footnote{https://scikit-learn.org/stable/modules/gen-erated/sklearn.svm.SVC.html}, which have been shown empirically to perform well with high dimensional data \citep{hsu2003practical}. We tune the hyper-parameters of the SVM models every 10 iterations (i.e. 100 labels requested) using the currently labelled dataset,  $\mathcal{L}$. In uncertainty sampling, the most uncertain examples are equivalent to those closest to the class separating hyper-plane in the context of an SVM \citep{tong2000support}. In the information density selection strategy, we use entropy to measure the ``informativeness'' and all parameters are set following \cite{settles2008analysis}. In QBC, we choose Linear-SVM models trained using bagging as committee members following \cite{mamitsuka1998query}. Since there is no general agreement in the literature on the appropriate committee size for QBC \citep{settles2009active}, we adopt a committee size 5 after some preliminary experiments. In EGAL, all parameters are set following the recommendations given in \cite{hu2010egal}, which are shown to perform well for text classification tasks. All experiments are run on the machine with GPU (NVIDIA GeForce GTX 1080 Ti) and CPU (Intel Core i7-8700K 3.70 GHz).

\subsection{Configuration of the Text Representation Techniques} \label{sec:text_configurations}

For the frequency based vector representations, BOW and LDA, we preprocess text data by converting to lowercase, removing stop words, and removing rare terms (for the whole dataset, word count less than 10 or document frequency less than 5). We set the number of topics to be used by LDA\footnote{https://radimrehurek.com/gensim/models/ldamodel.html} to 300 following \cite{singh2018improving}. 

In this experiment three different pre-trained word embedding representations are used: word2vec (the 300 dimension version trained using the google news corpus), Glove (the 300 dimension version trained using the wiki  gigaword corpus) and FastText (the 300 dimension version trained using the Wikipedia corpus) which are all trained with large online corpora. \footnote{https://radimrehurek.com/gensim/index.html} For all of the word embedding representations, we average the vectors of the words that appeared in the document to represent each document.

This experiment uses 4 of the most high-profile transformer-based language models: BERT (``bert-base-uncased''), XLNet (``xlnet-base-cased''), GPT-2 (``gpt2'') and Roberta (``roberta-base''). The codes beside the algorithms are the specific models used in the experiments which can be found on Github.\footnote{https://github.com/huggingface/transformers}  
Since all transformer-based models are configured to take as input a maximum of 512 tokens, we divided the long documents with $W$ words into $k = W/511$ fractions, which is then fed to the model to infer the representation of each fraction (each fraction has a ``[CLS]'' token in front of 511 tokens, so, 512 tokens in total). The vector of each fraction is the average embeddings of words in that fraction and the representation of the whole text sequence is the mean of all $k$ fraction vectors. It should be noted that we do not use any label information for fine-tuning any model to ensure fair comparisons. A summary of the dimensionality of each representation is given in Table \ref{tab:datasets}.

 \begin{table*}[h!]
\caption{Statistics of 8 datasets used on our experiments, the number of instances of each class and the length of the different representations generated (the representations generated using all of the trasnsformer-based models have the same dimensionality).}
\label{tab:datasets}

\small
    \centering
\begin{tabular}{l|rr|c r c c c c } \hline 
      \multicolumn{1}{c|}{} &
      \multicolumn{2}{c|}{\# of Instances}& \multicolumn{6}{c}{Representation Dimensionality} \\ 
     {\fontseries{b}\selectfont Dataset} & {\fontseries{b}\selectfont positives}      & {\fontseries{b}\selectfont negatives} & { \fontseries{b}\selectfont   LDA} & {\fontseries{b}\selectfont TF-IDF} & {\fontseries{b}\selectfont FastText} & 
     {     \fontseries{b}\selectfont   Glove} & {\fontseries{b}\selectfont  word2vec} & {\fontseries{b}\selectfont Transformer-based}  \\   \hline
MR         & 1,000          & 1,000          & 300 & 6,181& 300 & 300 & 300  & 768  \\
MDCR      &     4,000          & 3,566          & 300 & 4,165& 300 & 300 & 300  & 768  \\
BAG        &    1,675          & 1,552          & 300 & 4,936& 300 & 300 & 300  & 768  \\
G2013     &       843          & 1,292          & 300 & 5,345& 300 & 300 & 300  & 768  \\ \hline

ACR       &    1,335          & 736         &300 & 403& 300 & 300 & 300  & 768  \\

MRS     &      5,000          & 5,000         & 300 & 1,868& 300 & 300 & 300  & 768 \\ 
AGN     &      1,000          & 1,000         & 300 & 723& 300 & 300 & 300  & 768  \\
DBP     &      1,000          & 1,000         & 300 & 552& 300 & 300 & 300  & 768  \\
\hline

\end{tabular}

\end{table*}

\begin{table*}[h!]
\caption{The number of parameters and inference time cost by various pre-trained transformer-based language models (second). The best performing model is highlighted and the second to last row denotes the average ranking of each model. The last row shows the number of parameters of each model which implies the memory used by each model.}
\label{tab:gpu_inference_time}
\resizebox{\textwidth}{!}{%
\begin{tabular}{l|r|r|r|r|r|r}  \hline 
 & BERT & GPT-2 & XLNet & DistilBert & Albert & Roberta \\ \hline 
\textbf{MR} & 152.51$\pm$6.91(2.0) & 192.08$\pm$12.86(4.0) & 279.76$\pm$15.08(6.0) & \textbf{81.64$\pm$2.28(1.0)} & 193.07$\pm$8.64(5.0) & 165.39$\pm$12.33(3.0) \\
\textbf{MRS} & 367.3$\pm$22.9(2.0) & 418.4$\pm$24.9(5.0) & 475.16$\pm$20.44(6.0) & \textbf{193.12$\pm$13.48(1.0)} & 391.52$\pm$24.99(4.0) & 373.29$\pm$22.7(3.0) \\
\textbf{ACR} & 73.75$\pm$4.43(3.0) & 81.35$\pm$4.8(5.0) & 91.11$\pm$3.1(6.0) & \textbf{37.65$\pm$1.78(1.0)} & 76.96$\pm$4.03(4.0) & 73.46$\pm$4.0(2.0) \\
\textbf{BAG} & 191.49$\pm$10.53(2.0) & 237.89$\pm$9.79(5.0) & 286.5$\pm$8.28(6.0) & \textbf{100.94$\pm$3.47(1.0)} & 222.37$\pm$8.36(4.0) & 196.25$\pm$9.35(3.0) \\
\textbf{MDCR} & 295.56$\pm$14.2(2.0) & 349.03$\pm$16.2(5.0) & 411.43$\pm$19.39(6.0) & \textbf{161.61$\pm$7.04(1.0)} & 317.86$\pm$14.0(4.0) & 296.79$\pm$15.58(3.0) \\
\textbf{G2013} & 161.75$\pm$11.73(2.0) & 209.39$\pm$12.67(5.0) & 258.95$\pm$11.03(6.0) & \textbf{83.32$\pm$4.41(1.0)} & 193.41$\pm$10.05(4.0) & 173.87$\pm$10.54(3.0) \\
\textbf{AGN} & 71.52$\pm$6.21(3.0) & 78.97$\pm$6.87(5.0) & 92.64$\pm$7.45(6.0) & \textbf{36.99$\pm$3.69(1.0)} & 72.23$\pm$5.86(4.0) & 69.88$\pm$5.59(2.0) \\
\textbf{DBP} & 66.67$\pm$3.49(2.0) & 76.46$\pm$4.7(5.0) & 89.69$\pm$6.51(6.0) & \textbf{34.69$\pm$2.54(1.0)} & 69.68$\pm$4.46(4.0) & 67.33$\pm$4.01(3.0) \\ \hline 
\textbf{Rank} & 2.25 & 4.88 & 6.00 & \textbf{1.00} & 4.12 & 2.75 \\ \hline
\textbf{\#Params} & 109,482,240 & 124,439,808 & 116,718,336 & 66,362,880 & \textbf{11,683,584} & 124,645,632 \\ \hline  
\end{tabular}%
}
\end{table*}

\subsection{Performance Measures} \label{sec:metric}

As we are interested in the ability of an active learning process to fully label a dataset we use the \emph{accuracy+} performance measure, which has been previously used by \cite{hu2008sweetening,hu2010egal}. This measures the performance of the full active learning system including human annotators which is prevalent in active learning community \citep{wallace2010active,hashimoto2016topic}. It can be expressed as:

\begin{equation}
accuracy+ = \frac{TP^{H}+TN^{H}+TP^{M}+TN^{M}}{N}
\label{eq:acc}
\end{equation}

\noindent where $N$ is the total number of instances in a dataset  and superscripts $H$ and $M$ express human annotator and machine generated labels respectively. $TP$ and $TN$ denote the number of \emph{true positives} and \emph{true negatives} respectively. Intuitively, this metric computes the fraction of correctly labelled instances which are predicted by the oracles as well as a trained classifier. We presume that a human annotator never makes mistakes. 

We also report the \emph{area under the learning curve} (AULC) score based on accuracy+ to measure the overall performance of the active learning process when different representations are used. As we can see in Figure \ref{fig:MDCR_random}, each line represents a learning curve of a representation technique. X-axis represents the number of documents that have been manually annotated and Y-axis denotes accuracy+ and the AULC score is the area under each curve. For example, the area shaded yellow is the AULC score of learning curve of LDA representation. In this work, all AULC scores are computed using the trapezoidal rule and normalized by the maximum possible area, to bound the value between 0 and 1.

\subsection{Datasets} \label{sec:dataset}

We evaluate the performance of active learning systems using 8 fully-labelled datasets. Four of these datasets are based on long text sequences: \emph{Movie Review (MR)} \citep{pang2004sentimental},\footnote{\textbf{MR and MRS are available at:} http://www.cs.cornell.edu/people/pabo/movie-review-data/} \emph{Multi-domain Customer Review (MDCR)} \citep{blitzer2007biographies},\footnote{https://www.cs.jhu.edu/\textasciitilde mdredze/datasets/sentiment/index2.html} \emph{Blog Author Gender (BAG)} \citep{mukherjee2010improving}\footnote{\textbf{BAG and ACR are available at:} https://www.cs.uic.edu/\textasciitilde liub/FBS/sentiment-analysis.html} and \emph{Guardian2013 (G2013)} \citep{belford2018stability}. Four are based on sentences: \emph{Additional Customer Review (ACR)} \citep{ding2008holistic}, \emph{Movie Review Subjectivity (MRS)} \citep{pang2004sentimental}, \emph{Ag news (AGN)}\footnote{\textbf{AGN and DBP are available at :} https://skymind.ai/wiki/open-datasets} and \emph{Dbpedia (DBP)} \citep{zhang2015character}. Table \ref{tab:datasets} provides summary statistics describing each dataset.


\renewcommand{\arraystretch}{1.1}
\begin{table*}

\centering
\caption{The summary of AULC scores, which are computed by trapezoidal rule and normalized by the maximum possible area, on each dataset regarding the different combinations of text representations and selection strategies. The best performance of each dataset is highlighted and the last column denotes the average ranking of each method where the smaller number suggests a higher rank.}
\label{tab:AUC_summary_small}

\resizebox{1.0\textwidth}{!}{%

\begin{tabular}{l|r|r|r|r|r|r|r|r|r|r}
\hline
\textbf{Rep} & \textbf{Strategy} & \textbf{MR} & \textbf{MDCR} & \textbf{BAG} & \textbf{G2013} & \textbf{ACR} & \textbf{MRS} & \textbf{AGN} & \textbf{DBP} & \textbf{Rank} \\\hline
\multirow{1}{*}{BERT} 
 & uncertainty & 0.891$\pm$0.005(15.0) & 0.819$\pm$0.006(14.0) & 0.738$\pm$0.006(4.0) & 0.980$\pm$0.001(9.0) & 0.862$\pm$0.003(10.0) & 0.932$\pm$0.000(4.0) & 0.991$\pm$0.000(8.5) & \textbf{0.995$\pm$0.001(1.0)} & 8.19 \\
\hline
\multirow{1}{*}{XLNet} 
 & uncertainty & \textbf{0.952$\pm$0.004(1.0)} & 0.865$\pm$0.005(4.0) & 0.744$\pm$0.005(2.0) & \textbf{0.985$\pm$0.001(1.0)} & 0.886$\pm$0.007(5.0) & 0.917$\pm$0.001(11.0) & 0.994$\pm$0.000(3.0) & 0.994$\pm$0.001(4.0) & 3.88 \\\hline
\multirow{1}{*}{GPT-2} 
 & uncertainty & 0.911$\pm$0.004(8.0) & 0.843$\pm$0.003(9.0) & 0.739$\pm$0.006(3.0) & 0.984$\pm$0.001(2.0) & 0.875$\pm$0.003(7.0) & 0.930$\pm$0.001(6.0) & \textbf{0.995$\pm$0.000(1.0)} & 0.988$\pm$0.001(12.0) & 6.00 \\\hline
\multirow{1}{*}{Roberta}
 & uncertainty & 0.950$\pm$0.002(3.0) & \textbf{0.879$\pm$0.004(1.0)} & \textbf{0.747$\pm$0.007(1.0)} & 0.983$\pm$0.002(4.0) & \textbf{0.903$\pm$0.002(1.0)} & \textbf{0.949$\pm$0.001(1.0)} & 0.993$\pm$0.000(4.0) & 0.995$\pm$0.000(2.0) & \textbf{2.12} \\\hline
\multirow{1}{*}{Glove}
 & QBC & 0.845$\pm$0.003(30.0) & 0.715$\pm$0.008(30.0) & 0.701$\pm$0.018(28.5) & 0.977$\pm$0.001(16.0) & 0.787$\pm$0.009(23.0) & 0.900$\pm$0.003(21.0) & 0.971$\pm$0.001(30.0) & 0.978$\pm$0.002(24.0) & 25.31 \\\hline
\multirow{1}{*}{word2vec} 
 & QBC & 0.872$\pm$0.003(18.0) & 0.733$\pm$0.007(22.0) & 0.706$\pm$0.012(23.0) & 0.975$\pm$0.003(19.5) & 0.792$\pm$0.006(21.0) & 0.893$\pm$0.002(26.0) & 0.989$\pm$0.001(13.0) & 0.979$\pm$0.002(23.0) & 20.69 \\\hline
\multirow{1}{*}{FastText}
 & QBC & 0.857$\pm$0.005(24.0) & 0.724$\pm$0.007(26.5) & 0.706$\pm$0.013(22.0) & 0.976$\pm$0.001(18.0) & 0.788$\pm$0.007(22.0) & 0.904$\pm$0.002(17.0) & 0.976$\pm$0.002(24.0) & 0.993$\pm$0.002(8.5) & 20.25 \\\hline
\multirow{1}{*}{TF-IDF}
 & QBC & 0.854$\pm$0.012(26.0) & 0.713$\pm$0.007(31.0) & 0.676$\pm$0.007(37.0) & 0.965$\pm$0.002(24.0) & 0.778$\pm$0.005(24.0) & 0.812$\pm$0.005(35.0) & 0.932$\pm$0.004(36.0) & 0.952$\pm$0.007(33.0) & 30.75 \\\hline
\multirow{1}{*}{LDA}
 & QBC & 0.770$\pm$0.009(44.0) & 0.607$\pm$0.009(43.0) & 0.675$\pm$0.008(38.0) & 0.917$\pm$0.008(42.0) & 0.763$\pm$0.008(31.0) & 0.689$\pm$0.006(41.0) & 0.901$\pm$0.005(39.0) & 0.905$\pm$0.006(41.0) & 39.88 \\\hline
\end{tabular}%

}

\end{table*}

\begin{landscape}
\renewcommand{\arraystretch}{1.1}
\begin{table}

\centering
\caption{The summary of AULC scores, which are computed by trapezoidal rule and normalized by the maximum possible area, on each dataset regarding the different combinations of text representations and selection strategies. The best performance of each dataset is highlighted and the last column denotes the average ranking of each method where the smaller number suggests a higher rank.}
\label{tab:AUC_summary}

\resizebox{1.2\textwidth}{!}{%

\begin{tabular}{l|r|r|r|r|r|r|r|r|r|r}
\hline
\textbf{Rep} & \textbf{Strategy} & \textbf{MR} & \textbf{MDCR} & \textbf{BAG} & \textbf{G2013} & \textbf{ACR} & \textbf{MRS} & \textbf{AGN} & \textbf{DBP} & \textbf{Rank} \\\hline
\multirow{5}{*}{BERT} & random & 0.856$\pm$0.006(25.0) & 0.800$\pm$0.003(17.0) & 0.722$\pm$0.005(17.0) & 0.947$\pm$0.003(36.5) & 0.816$\pm$0.006(18.0) & 0.913$\pm$0.002(12.5) & 0.971$\pm$0.002(29.0) & 0.986$\pm$0.001(15.0) & 21.25 \\
 & uncertainty & 0.891$\pm$0.005(15.0) & 0.819$\pm$0.006(14.0) & 0.738$\pm$0.006(4.0) & 0.980$\pm$0.001(9.0) & 0.862$\pm$0.003(10.0) & 0.932$\pm$0.000(4.0) & 0.991$\pm$0.000(8.5) & \textbf{0.995$\pm$0.001(1.0)} & 8.19 \\
 & EGAL & 0.841$\pm$0.013(32.0) & 0.738$\pm$0.010(20.0) & 0.693$\pm$0.005(32.0) & 0.949$\pm$0.003(33.0) & 0.818$\pm$0.005(17.0) & 0.897$\pm$0.010(23.0) & 0.973$\pm$0.001(27.0) & 0.980$\pm$0.004(22.0) & 25.75 \\
 & ID & 0.893$\pm$0.005(14.0) & 0.810$\pm$0.006(16.0) & 0.733$\pm$0.012(9.0) & 0.978$\pm$0.001(11.5) & 0.846$\pm$0.006(14.0) & 0.926$\pm$0.003(9.0) & 0.991$\pm$0.000(8.5) & 0.994$\pm$0.001(4.0) & 10.75 \\
 & QBC & 0.887$\pm$0.004(16.0) & 0.813$\pm$0.005(15.0) & 0.724$\pm$0.014(15.0) & 0.973$\pm$0.002(22.0) & 0.858$\pm$0.005(11.0) & 0.928$\pm$0.003(7.0) & 0.988$\pm$0.002(14.0) & 0.994$\pm$0.001(4.0) & 13.00 \\\hline
\multirow{5}{*}{XLNet} & random & 0.913$\pm$0.003(7.0) & 0.852$\pm$0.004(8.0) & 0.732$\pm$0.002(12.0) & 0.959$\pm$0.002(26.0) & 0.839$\pm$0.007(15.0) & 0.901$\pm$0.003(19.0) & 0.980$\pm$0.001(18.5) & 0.983$\pm$0.001(19.0) & 15.56 \\
 & uncertainty & \textbf{0.952$\pm$0.004(1.0)} & 0.865$\pm$0.005(4.0) & 0.744$\pm$0.005(2.0) & \textbf{0.985$\pm$0.001(1.0)} & 0.886$\pm$0.007(5.0) & 0.917$\pm$0.001(11.0) & 0.994$\pm$0.000(3.0) & 0.994$\pm$0.001(4.0) & 3.88 \\
 & EGAL & 0.899$\pm$0.011(13.0) & 0.789$\pm$0.025(18.0) & 0.719$\pm$0.007(18.0) & 0.963$\pm$0.001(25.0) & 0.821$\pm$0.009(16.0) & 0.882$\pm$0.010(32.0) & 0.979$\pm$0.001(20.0) & 0.977$\pm$0.003(25.0) & 20.88 \\
 & ID & 0.950$\pm$0.003(2.0) & 0.859$\pm$0.008(6.0) & 0.733$\pm$0.007(10.0) & 0.983$\pm$0.003(3.0) & 0.889$\pm$0.006(4.0) & 0.911$\pm$0.005(14.0) & 0.991$\pm$0.002(7.0) & 0.994$\pm$0.000(6.5) & 6.56 \\
 & QBC & 0.943$\pm$0.005(6.0) & 0.862$\pm$0.005(5.0) & 0.734$\pm$0.009(8.0) & 0.979$\pm$0.002(10.0) & 0.884$\pm$0.003(6.0) & 0.913$\pm$0.002(12.5) & 0.990$\pm$0.001(11.5) & 0.992$\pm$0.002(10.0) & 8.62 \\\hline
\multirow{5}{*}{GPT-2} & random & 0.875$\pm$0.006(17.0) & 0.826$\pm$0.004(13.0) & 0.727$\pm$0.004(14.0) & 0.955$\pm$0.003(27.0) & 0.815$\pm$0.005(19.0) & 0.909$\pm$0.002(15.0) & 0.983$\pm$0.001(16.0) & 0.967$\pm$0.002(31.0) & 19.00 \\
 & uncertainty & 0.911$\pm$0.004(8.0) & 0.843$\pm$0.003(9.0) & 0.739$\pm$0.006(3.0) & 0.984$\pm$0.001(2.0) & 0.875$\pm$0.003(7.0) & 0.930$\pm$0.001(6.0) & \textbf{0.995$\pm$0.000(1.0)} & 0.988$\pm$0.001(12.0) & 6.00 \\
 & EGAL & 0.865$\pm$0.003(22.0) & 0.749$\pm$0.021(19.0) & 0.723$\pm$0.006(16.0) & 0.954$\pm$0.001(29.0) & 0.812$\pm$0.006(20.0) & 0.904$\pm$0.005(16.0) & 0.982$\pm$0.001(17.0) & 0.970$\pm$0.003(29.0) & 21.00 \\
 & ID & 0.908$\pm$0.004(10.0) & 0.838$\pm$0.004(10.0) & 0.734$\pm$0.010(7.0) & 0.983$\pm$0.001(5.5) & 0.862$\pm$0.008(9.0) & 0.927$\pm$0.002(8.0) & 0.994$\pm$0.001(2.0) & 0.987$\pm$0.001(14.0) & 8.19 \\
 & QBC & 0.906$\pm$0.003(11.0) & 0.834$\pm$0.005(12.0) & 0.730$\pm$0.008(13.0) & 0.977$\pm$0.002(15.0) & 0.867$\pm$0.005(8.0) & 0.925$\pm$0.006(10.0) & 0.992$\pm$0.001(5.5) & 0.984$\pm$0.002(17.0) & 11.44 \\\hline
\multirow{5}{*}{Roberta} & random & 0.910$\pm$0.003(9.0) & 0.858$\pm$0.005(7.0) & 0.732$\pm$0.003(11.0) & 0.954$\pm$0.003(28.0) & 0.852$\pm$0.004(12.0) & 0.931$\pm$0.001(5.0) & 0.976$\pm$0.001(25.0) & 0.984$\pm$0.001(18.0) & 14.38 \\
 & uncertainty & 0.950$\pm$0.002(3.0) & \textbf{0.879$\pm$0.004(1.0)} & \textbf{0.747$\pm$0.007(1.0)} & 0.983$\pm$0.002(4.0) & \textbf{0.903$\pm$0.002(1.0)} & \textbf{0.949$\pm$0.001(1.0)} & 0.993$\pm$0.000(4.0) & 0.995$\pm$0.000(2.0) & \textbf{2.12} \\
 & EGAL & 0.905$\pm$0.006(12.0) & 0.835$\pm$0.011(11.0) & 0.714$\pm$0.005(19.0) & 0.953$\pm$0.003(30.0) & 0.852$\pm$0.003(13.0) & 0.903$\pm$0.009(18.0) & 0.977$\pm$0.002(21.0) & 0.985$\pm$0.003(16.0) & 17.50 \\
 & ID & 0.949$\pm$0.003(4.0) & 0.873$\pm$0.005(2.0) & 0.736$\pm$0.006(5.0) & 0.983$\pm$0.001(5.5) & 0.894$\pm$0.003(2.0) & 0.940$\pm$0.002(3.0) & 0.992$\pm$0.001(5.5) & 0.994$\pm$0.000(6.5) & 4.19 \\
 & QBC & 0.945$\pm$0.002(5.0) & 0.869$\pm$0.004(3.0) & 0.735$\pm$0.008(6.0) & 0.978$\pm$0.001(11.5) & 0.892$\pm$0.006(3.0) & 0.946$\pm$0.002(2.0) & 0.990$\pm$0.001(11.5) & 0.993$\pm$0.002(8.5) & 6.31 \\\hline
\multirow{5}{*}{Glove} & random & 0.812$\pm$0.002(38.0) & 0.712$\pm$0.009(32.0) & 0.705$\pm$0.004(25.0) & 0.950$\pm$0.002(31.5) & 0.741$\pm$0.006(38.0) & 0.887$\pm$0.002(30.0) & 0.936$\pm$0.002(35.0) & 0.946$\pm$0.007(34.0) & 32.94 \\
 & uncertainty & 0.840$\pm$0.004(33.0) & 0.710$\pm$0.021(33.0) & 0.708$\pm$0.011(21.0) & 0.977$\pm$0.004(14.0) & 0.770$\pm$0.021(27.0) & 0.900$\pm$0.010(20.0) & 0.970$\pm$0.004(31.0) & 0.971$\pm$0.007(28.0) & 25.88 \\
 & EGAL & 0.808$\pm$0.006(39.0) & 0.615$\pm$0.018(40.0) & 0.691$\pm$0.009(34.0) & 0.948$\pm$0.003(34.0) & 0.719$\pm$0.006(45.0) & 0.793$\pm$0.038(40.0) & 0.939$\pm$0.002(34.0) & 0.944$\pm$0.008(35.0) & 37.62 \\
 & ID & 0.845$\pm$0.008(29.0) & 0.704$\pm$0.015(35.0) & 0.692$\pm$0.014(33.0) & 0.977$\pm$0.005(13.0) & 0.765$\pm$0.009(29.0) & 0.891$\pm$0.002(28.0) & 0.971$\pm$0.004(28.0) & 0.969$\pm$0.008(30.0) & 28.12 \\
 & QBC & 0.845$\pm$0.003(30.0) & 0.715$\pm$0.008(30.0) & 0.701$\pm$0.018(28.5) & 0.977$\pm$0.001(16.0) & 0.787$\pm$0.009(23.0) & 0.900$\pm$0.003(21.0) & 0.971$\pm$0.001(30.0) & 0.978$\pm$0.002(24.0) & 25.31 \\\hline
\multirow{5}{*}{word2vec} & random & 0.842$\pm$0.005(31.0) & 0.731$\pm$0.004(23.0) & 0.702$\pm$0.009(27.0) & 0.947$\pm$0.004(35.0) & 0.749$\pm$0.007(35.0) & 0.885$\pm$0.004(31.0) & 0.975$\pm$0.001(26.0) & 0.953$\pm$0.009(32.0) & 30.00 \\
 & uncertainty & 0.865$\pm$0.008(21.0) & 0.734$\pm$0.014(21.0) & 0.713$\pm$0.012(20.0) & 0.975$\pm$0.003(19.5) & 0.767$\pm$0.027(28.0) & 0.894$\pm$0.005(25.0) & 0.987$\pm$0.005(15.0) & 0.974$\pm$0.011(26.0) & 21.94 \\
 & EGAL & 0.829$\pm$0.006(35.0) & 0.654$\pm$0.024(37.0) & 0.699$\pm$0.011(30.0) & 0.943$\pm$0.003(39.0) & 0.725$\pm$0.007(43.0) & 0.829$\pm$0.022(34.0) & 0.976$\pm$0.003(23.0) & 0.936$\pm$0.010(37.0) & 34.75 \\
 & ID & 0.866$\pm$0.010(20.0) & 0.727$\pm$0.009(25.0) & 0.703$\pm$0.013(26.0) & 0.976$\pm$0.002(17.0) & 0.763$\pm$0.015(30.0) & 0.891$\pm$0.004(27.0) & 0.990$\pm$0.002(10.0) & 0.973$\pm$0.008(27.0) & 22.75 \\
 & QBC & 0.872$\pm$0.003(18.0) & 0.733$\pm$0.007(22.0) & 0.706$\pm$0.012(23.0) & 0.975$\pm$0.003(19.5) & 0.792$\pm$0.006(21.0) & 0.893$\pm$0.002(26.0) & 0.989$\pm$0.001(13.0) & 0.979$\pm$0.002(23.0) & 20.69 \\\hline
\multirow{5}{*}{FastText} & random & 0.821$\pm$0.006(36.0) & 0.724$\pm$0.007(26.5) & 0.705$\pm$0.006(24.0) & 0.950$\pm$0.002(31.5) & 0.740$\pm$0.011(39.5) & 0.888$\pm$0.003(29.0) & 0.950$\pm$0.002(33.0) & 0.982$\pm$0.004(21.0) & 30.06 \\
 & uncertainty & 0.853$\pm$0.008(27.0) & 0.728$\pm$0.014(24.0) & 0.701$\pm$0.018(28.5) & 0.980$\pm$0.003(8.0) & 0.776$\pm$0.018(25.0) & 0.894$\pm$0.011(24.0) & 0.976$\pm$0.008(22.0) & 0.988$\pm$0.006(11.0) & 21.19 \\
 & EGAL & 0.814$\pm$0.005(37.0) & 0.647$\pm$0.041(38.0) & 0.656$\pm$0.018(43.0) & 0.947$\pm$0.003(36.5) & 0.737$\pm$0.009(41.0) & 0.859$\pm$0.031(33.0) & 0.957$\pm$0.001(32.0) & 0.982$\pm$0.006(20.0) & 35.06 \\
 & ID & 0.852$\pm$0.008(28.0) & 0.720$\pm$0.008(28.0) & 0.697$\pm$0.015(31.0) & 0.981$\pm$0.002(7.0) & 0.776$\pm$0.012(26.0) & 0.898$\pm$0.006(22.0) & 0.980$\pm$0.001(18.5) & 0.987$\pm$0.006(13.0) & 21.69 \\
 & QBC & 0.857$\pm$0.005(24.0) & 0.724$\pm$0.007(26.5) & 0.706$\pm$0.013(22.0) & 0.976$\pm$0.001(18.0) & 0.788$\pm$0.007(22.0) & 0.904$\pm$0.002(17.0) & 0.976$\pm$0.002(24.0) & 0.993$\pm$0.002(8.5) & 20.25 \\\hline
\multirow{5}{*}{TF-IDF} & random & 0.837$\pm$0.003(34.0) & 0.708$\pm$0.012(34.0) & 0.666$\pm$0.006(42.0) & 0.945$\pm$0.002(38.0) & 0.740$\pm$0.011(39.5) & 0.808$\pm$0.007(37.0) & 0.887$\pm$0.011(41.0) & 0.912$\pm$0.014(40.0) & 38.19 \\
 & uncertainty & 0.871$\pm$0.005(19.0) & 0.719$\pm$0.009(29.0) & 0.684$\pm$0.006(35.0) & 0.975$\pm$0.001(21.0) & 0.758$\pm$0.017(32.0) & 0.807$\pm$0.016(38.0) & 0.919$\pm$0.017(37.0) & 0.943$\pm$0.021(36.0) & 30.88 \\
 & EGAL & 0.800$\pm$0.008(40.0) & 0.642$\pm$0.043(39.0) & 0.637$\pm$0.009(45.0) & 0.942$\pm$0.005(40.0) & 0.745$\pm$0.009(36.0) & 0.809$\pm$0.008(36.0) & 0.895$\pm$0.012(40.0) & 0.912$\pm$0.016(39.0) & 39.38 \\
 & ID & 0.862$\pm$0.004(23.0) & 0.696$\pm$0.004(36.0) & 0.683$\pm$0.003(36.0) & 0.971$\pm$0.002(23.0) & 0.744$\pm$0.013(37.0) & 0.803$\pm$0.016(39.0) & 0.915$\pm$0.010(38.0) & 0.926$\pm$0.018(38.0) & 33.75 \\
 & QBC & 0.854$\pm$0.012(26.0) & 0.713$\pm$0.007(31.0) & 0.676$\pm$0.007(37.0) & 0.965$\pm$0.002(24.0) & 0.778$\pm$0.005(24.0) & 0.812$\pm$0.005(35.0) & 0.932$\pm$0.004(36.0) & 0.952$\pm$0.007(33.0) & 30.75 \\\hline
\multirow{5}{*}{LDA} & random & 0.772$\pm$0.006(43.0) & 0.611$\pm$0.006(42.0) & 0.669$\pm$0.005(41.0) & 0.884$\pm$0.012(44.0) & 0.733$\pm$0.006(42.0) & 0.680$\pm$0.003(42.0) & 0.854$\pm$0.004(44.0) & 0.858$\pm$0.006(44.0) & 42.75 \\
 & uncertainty & 0.791$\pm$0.006(42.0) & 0.613$\pm$0.010(41.0) & 0.673$\pm$0.015(39.0) & 0.922$\pm$0.012(41.0) & 0.754$\pm$0.011(33.0) & 0.671$\pm$0.013(43.0) & 0.877$\pm$0.017(42.0) & 0.881$\pm$0.011(42.0) & 40.38 \\
 & EGAL & 0.760$\pm$0.007(45.0) & 0.600$\pm$0.008(45.0) & 0.648$\pm$0.009(44.0) & 0.858$\pm$0.014(45.0) & 0.720$\pm$0.008(44.0) & 0.611$\pm$0.015(45.0) & 0.835$\pm$0.006(45.0) & 0.805$\pm$0.018(45.0) & 44.75 \\
 & ID & 0.796$\pm$0.005(41.0) & 0.606$\pm$0.006(44.0) & 0.672$\pm$0.005(40.0) & 0.914$\pm$0.010(43.0) & 0.750$\pm$0.010(34.0) & 0.620$\pm$0.008(44.0) & 0.862$\pm$0.010(43.0) & 0.863$\pm$0.012(43.0) & 41.50 \\
 & QBC & 0.770$\pm$0.009(44.0) & 0.607$\pm$0.009(43.0) & 0.675$\pm$0.008(38.0) & 0.917$\pm$0.008(42.0) & 0.763$\pm$0.008(31.0) & 0.689$\pm$0.006(41.0) & 0.901$\pm$0.005(39.0) & 0.905$\pm$0.006(41.0) & 39.88 \\\hline
\end{tabular}%

}

\end{table}
\end{landscape}

\begin{landscape}
\renewcommand{\arraystretch}{1.1}
\begin{table}[h!]

\centering
\caption{The summary of AULC scores, which are computed by trapezoidal rule and normalized by the maximum possible area, on each dataset regarding the different combinations of text representations and selection strategies. The best performance of each dataset is highlighted and the last column denotes the average ranking of each method where the smaller number suggests a higher rank.}
\label{tab:AUC_CLS_summary}

\resizebox{1.2\textwidth}{!}{%

\begin{tabular}{l|r|r|r|r|r|r|r|r|r|r}
\hline
\textbf{Rep} & \textbf{Strategy} & \textbf{MR} & \textbf{MDCR} & \textbf{BAG} & \textbf{G2013} & \textbf{ACR} & \textbf{MRS} & \textbf{AGN} & \textbf{DBP} & \textbf{Rank} \\\hline
\multirow{5}{*}{BERT} & random & 0.856$\pm$0.006(18.0) & 0.800$\pm$0.003(21.0) & 0.722$\pm$0.005(19.0) & 0.947$\pm$0.003(26.0) & 0.816$\pm$0.006(27.0) & 0.913$\pm$0.002(21.0) & 0.971$\pm$0.002(27.0) & 0.986$\pm$0.001(23.0) & 22.75 \\
 & uncertainty & 0.891$\pm$0.005(12.0) & 0.819$\pm$0.006(10.0) & 0.738$\pm$0.006(4.0) & 0.980$\pm$0.001(6.0) & 0.862$\pm$0.003(12.0) & 0.932$\pm$0.000(12.0) & 0.991$\pm$0.000(11.5) & 0.995$\pm$0.001(6.0) & 9.19 \\
 & EGAL & 0.841$\pm$0.013(26.0) & 0.738$\pm$0.010(29.0) & 0.693$\pm$0.005(30.0) & 0.949$\pm$0.003(24.0) & 0.818$\pm$0.005(26.0) & 0.897$\pm$0.010(27.0) & 0.973$\pm$0.001(26.0) & 0.980$\pm$0.004(26.0) & 26.75 \\
 & ID & 0.893$\pm$0.005(10.0) & 0.810$\pm$0.006(15.0) & 0.733$\pm$0.012(11.0) & 0.978$\pm$0.001(9.0) & 0.846$\pm$0.006(18.0) & 0.926$\pm$0.003(16.0) & 0.991$\pm$0.000(11.5) & 0.994$\pm$0.001(9.5) & 12.50 \\
 & QBC & 0.887$\pm$0.004(15.0) & 0.813$\pm$0.005(13.0) & 0.724$\pm$0.014(18.0) & 0.973$\pm$0.002(13.5) & 0.858$\pm$0.005(13.0) & 0.928$\pm$0.003(15.0) & 0.988$\pm$0.002(14.0) & 0.994$\pm$0.001(9.5) & 13.88 \\\hline
\multirow{5}{*}{Bert\_CLS} & random & 0.821$\pm$0.004(27.0) & 0.770$\pm$0.004(25.0) & 0.702$\pm$0.006(28.0) & 0.935$\pm$0.004(29.0) & 0.802$\pm$0.002(29.0) & 0.891$\pm$0.003(28.5) & 0.951$\pm$0.002(30.0) & 0.974$\pm$0.003(28.0) & 28.06 \\
 & uncertainty & 0.854$\pm$0.005(19.0) & 0.787$\pm$0.006(22.0) & 0.712$\pm$0.008(23.0) & 0.971$\pm$0.001(16.0) & 0.841$\pm$0.007(20.0) & 0.913$\pm$0.001(22.0) & 0.983$\pm$0.001(17.0) & 0.989$\pm$0.001(20.0) & 19.88 \\
 & EGAL & 0.802$\pm$0.005(29.0) & 0.736$\pm$0.040(30.0) & 0.678$\pm$0.011(33.0) & 0.934$\pm$0.004(30.0) & 0.774$\pm$0.002(30.0) & 0.891$\pm$0.003(28.5) & 0.954$\pm$0.003(29.0) & 0.965$\pm$0.005(30.0) & 29.94 \\
 & id & 0.850$\pm$0.003(21.0) & 0.768$\pm$0.005(26.0) & 0.705$\pm$0.010(25.0) & 0.967$\pm$0.001(17.0) & 0.835$\pm$0.005(21.0) & 0.905$\pm$0.002(24.0) & 0.982$\pm$0.002(18.0) & 0.988$\pm$0.001(21.0) & 21.62 \\
 & QBC & 0.846$\pm$0.006(23.0) & 0.774$\pm$0.006(24.0) & 0.710$\pm$0.007(24.0) & 0.966$\pm$0.002(18.0) & 0.844$\pm$0.005(19.0) & 0.911$\pm$0.006(23.0) & 0.977$\pm$0.002(23.5) & 0.987$\pm$0.002(22.0) & 22.06 \\\hline
\multirow{5}{*}{Roberta} & random & 0.910$\pm$0.003(7.0) & 0.858$\pm$0.005(5.0) & 0.732$\pm$0.003(13.0) & 0.954$\pm$0.003(21.0) & 0.852$\pm$0.004(15.0) & 0.931$\pm$0.001(13.0) & 0.976$\pm$0.001(25.0) & 0.984$\pm$0.001(25.0) & 15.50 \\
 & uncertainty & \textbf{0.950$\pm$0.002(1.0)} & \textbf{0.879$\pm$0.004(1.0)} & 0.747$\pm$0.007(2.0) & 0.983$\pm$0.002(2.0) & \textbf{0.903$\pm$0.002(1.0)} & \textbf{0.949$\pm$0.001(1.0)} & 0.993$\pm$0.000(5.0) & 0.995$\pm$0.000(8.0) & \textbf{2.62} \\
 & EGAL & 0.905$\pm$0.006(8.0) & 0.835$\pm$0.011(9.0) & 0.714$\pm$0.005(21.0) & 0.953$\pm$0.003(22.0) & 0.852$\pm$0.003(16.0) & 0.903$\pm$0.009(26.0) & 0.977$\pm$0.002(23.5) & 0.985$\pm$0.003(24.0) & 18.69 \\
 & ID & 0.949$\pm$0.003(2.0) & 0.873$\pm$0.005(2.0) & 0.736$\pm$0.006(7.0) & 0.983$\pm$0.001(3.0) & 0.894$\pm$0.003(4.0) & 0.940$\pm$0.002(7.0) & 0.992$\pm$0.001(8.5) & 0.994$\pm$0.000(11.0) & 5.56 \\
 & QBC & 0.945$\pm$0.002(3.0) & 0.869$\pm$0.004(3.0) & 0.735$\pm$0.008(9.0) & 0.978$\pm$0.001(9.0) & 0.892$\pm$0.006(5.0) & 0.946$\pm$0.002(2.0) & 0.990$\pm$0.001(13.0) & 0.993$\pm$0.002(12.0) & 7.00 \\\hline
\multirow{5}{*}{Roberta\_CLS} & random & 0.892$\pm$0.006(11.0) & 0.839$\pm$0.006(8.0) & 0.733$\pm$0.003(12.0) & 0.956$\pm$0.003(19.5) & 0.850$\pm$0.005(17.0) & 0.921$\pm$0.001(18.0) & 0.981$\pm$0.001(21.0) & 0.980$\pm$0.003(27.0) & 16.69 \\
 & uncertainty & 0.933$\pm$0.002(4.0) & 0.860$\pm$0.005(4.0) & \textbf{0.749$\pm$0.007(1.0)} & \textbf{0.984$\pm$0.001(1.0)} & 0.895$\pm$0.006(3.0) & 0.942$\pm$0.002(5.0) & 0.992$\pm$0.001(8.5) & 0.992$\pm$0.002(13.5) & 5.00 \\
 & EGAL & 0.878$\pm$0.010(16.0) & 0.786$\pm$0.029(23.0) & 0.713$\pm$0.012(22.0) & 0.956$\pm$0.003(19.5) & 0.818$\pm$0.011(24.0) & 0.915$\pm$0.005(20.0) & 0.966$\pm$0.009(28.0) & 0.969$\pm$0.005(29.0) & 22.69 \\
 & ID & 0.931$\pm$0.005(5.0) & 0.857$\pm$0.007(6.0) & 0.735$\pm$0.010(8.0) & 0.981$\pm$0.003(4.0) & 0.896$\pm$0.006(2.0) & 0.936$\pm$0.003(10.0) & 0.992$\pm$0.001(8.5) & 0.992$\pm$0.002(13.5) & 7.12 \\
 & QBC & 0.928$\pm$0.002(6.0) & 0.855$\pm$0.005(7.0) & 0.738$\pm$0.008(3.0) & 0.978$\pm$0.001(9.0) & 0.892$\pm$0.002(6.0) & 0.938$\pm$0.002(9.0) & 0.988$\pm$0.001(15.0) & 0.991$\pm$0.001(16.0) & 8.88 \\\hline
\multirow{5}{*}{DistilBert} & random & 0.861$\pm$0.005(17.0) & 0.803$\pm$0.006(20.0) & 0.726$\pm$0.005(16.0) & 0.948$\pm$0.002(25.0) & 0.821$\pm$0.005(23.0) & 0.922$\pm$0.002(17.0) & 0.981$\pm$0.001(21.0) & 0.990$\pm$0.002(18.5) & 19.69 \\
 & uncertainty & 0.894$\pm$0.005(9.0) & 0.815$\pm$0.007(12.0) & 0.737$\pm$0.008(5.0) & 0.980$\pm$0.001(6.0) & 0.867$\pm$0.007(9.0) & 0.941$\pm$0.002(6.0) & \textbf{0.995$\pm$0.000(1.5)} & \textbf{0.996$\pm$0.001(2.0)} & 6.31 \\
 & EGAL & 0.842$\pm$0.011(25.0) & 0.751$\pm$0.006(28.0) & 0.704$\pm$0.009(27.0) & 0.951$\pm$0.002(23.0) & 0.818$\pm$0.006(25.0) & 0.890$\pm$0.011(30.0) & 0.983$\pm$0.002(16.0) & 0.990$\pm$0.002(18.5) & 24.06 \\
 & ID & 0.890$\pm$0.003(14.0) & 0.807$\pm$0.006(17.0) & 0.736$\pm$0.009(6.0) & 0.977$\pm$0.003(11.0) & 0.856$\pm$0.003(14.0) & 0.930$\pm$0.001(14.0) & 0.994$\pm$0.000(3.5) & 0.995$\pm$0.001(6.0) & 10.69 \\
 & QBC & 0.890$\pm$0.006(13.0) & 0.812$\pm$0.007(14.0) & 0.729$\pm$0.014(14.0) & 0.973$\pm$0.002(13.5) & 0.867$\pm$0.005(10.0) & 0.939$\pm$0.002(8.0) & 0.992$\pm$0.001(8.5) & 0.995$\pm$0.001(6.0) & 10.88 \\\hline
\multirow{5}{*}{DistilBert\_CLS} & random & 0.818$\pm$0.004(28.0) & 0.805$\pm$0.003(19.0) & 0.717$\pm$0.005(20.0) & 0.946$\pm$0.002(27.0) & 0.829$\pm$0.004(22.0) & 0.919$\pm$0.002(19.0) & 0.981$\pm$0.001(21.0) & 0.991$\pm$0.001(16.0) & 21.50 \\
 & uncertainty & 0.850$\pm$0.004(20.0) & 0.817$\pm$0.007(11.0) & 0.727$\pm$0.011(15.0) & 0.980$\pm$0.001(6.0) & 0.876$\pm$0.009(7.0) & 0.944$\pm$0.001(3.0) & \textbf{0.995$\pm$0.000(1.5)} & 0.996$\pm$0.000(4.0) & 8.44 \\
 & EGAL & 0.791$\pm$0.010(30.0) & 0.758$\pm$0.029(27.0) & 0.705$\pm$0.005(26.0) & 0.945$\pm$0.003(28.0) & 0.813$\pm$0.005(28.0) & 0.904$\pm$0.003(25.0) & 0.982$\pm$0.001(19.0) & 0.991$\pm$0.001(16.0) & 24.88 \\
 & ID & 0.846$\pm$0.003(24.0) & 0.807$\pm$0.005(18.0) & 0.734$\pm$0.005(10.0) & 0.977$\pm$0.001(12.0) & 0.865$\pm$0.012(11.0) & 0.934$\pm$0.002(11.0) & 0.994$\pm$0.000(3.5) & \textbf{0.996$\pm$0.001(2.0)} & 11.44 \\
 & QBC & 0.847$\pm$0.006(22.0) & 0.809$\pm$0.007(16.0) & 0.725$\pm$0.009(17.0) & 0.973$\pm$0.001(15.0) & 0.874$\pm$0.008(8.0) & 0.943$\pm$0.002(4.0) & 0.992$\pm$0.002(6.0) & \textbf{0.996$\pm$0.001(2.0)} & 11.25 \\\hline
\multirow{5}{*}{Albert} & random & 0.741$\pm$0.005(34.0) & 0.642$\pm$0.008(33.0) & 0.681$\pm$0.005(31.0) & 0.882$\pm$0.006(37.0) & 0.711$\pm$0.006(36.0) & 0.779$\pm$0.004(33.0) & 0.903$\pm$0.003(37.0) & 0.828$\pm$0.005(34.0) & 34.38 \\
 & uncertainty & 0.761$\pm$0.009(31.0) & 0.647$\pm$0.009(32.0) & 0.695$\pm$0.008(29.0) & 0.927$\pm$0.008(31.0) & 0.728$\pm$0.018(34.0) & 0.788$\pm$0.007(32.0) & 0.948$\pm$0.004(31.0) & 0.875$\pm$0.005(31.0) & 31.38 \\
 & EGAL & 0.735$\pm$0.007(35.0) & 0.611$\pm$0.011(35.0) & 0.668$\pm$0.007(37.0) & 0.871$\pm$0.012(38.0) & 0.694$\pm$0.009(39.0) & 0.747$\pm$0.015(38.0) & 0.891$\pm$0.006(38.0) & 0.811$\pm$0.008(36.0) & 37.00 \\
 & ID & 0.754$\pm$0.009(33.0) & 0.623$\pm$0.006(34.0) & 0.675$\pm$0.017(34.0) & 0.922$\pm$0.008(33.0) & 0.720$\pm$0.014(35.0) & 0.774$\pm$0.010(34.0) & 0.941$\pm$0.004(33.0) & 0.856$\pm$0.008(33.0) & 33.62 \\
 & QBC & 0.758$\pm$0.008(32.0) & 0.651$\pm$0.009(31.0) & 0.680$\pm$0.010(32.0) & 0.925$\pm$0.003(32.0) & 0.748$\pm$0.011(31.0) & 0.791$\pm$0.004(31.0) & 0.945$\pm$0.002(32.0) & 0.872$\pm$0.006(32.0) & 31.62 \\\hline
\multirow{5}{*}{Albert\_CLS} & random & 0.698$\pm$0.005(38.0) & 0.597$\pm$0.007(38.0) & 0.659$\pm$0.007(39.0) & 0.850$\pm$0.006(39.0) & 0.704$\pm$0.008(37.0) & 0.741$\pm$0.002(39.0) & 0.873$\pm$0.005(39.0) & 0.771$\pm$0.005(39.0) & 38.50 \\
 & uncertainty & 0.704$\pm$0.005(36.0) & 0.607$\pm$0.005(36.0) & 0.672$\pm$0.012(35.0) & 0.889$\pm$0.007(34.0) & 0.731$\pm$0.016(33.0) & 0.755$\pm$0.005(35.0) & 0.926$\pm$0.004(34.0) & 0.812$\pm$0.007(35.0) & 34.75 \\
 & EGAL & 0.688$\pm$0.005(40.0) & 0.559$\pm$0.009(40.0) & 0.635$\pm$0.010(40.0) & 0.823$\pm$0.018(40.0) & 0.686$\pm$0.015(40.0) & 0.706$\pm$0.018(40.0) & 0.868$\pm$0.004(40.0) & 0.766$\pm$0.009(40.0) & 40.00 \\
 & ID & 0.692$\pm$0.006(39.0) & 0.595$\pm$0.006(39.0) & 0.670$\pm$0.007(36.0) & 0.885$\pm$0.009(36.0) & 0.701$\pm$0.021(38.0) & 0.750$\pm$0.008(37.0) & 0.917$\pm$0.007(36.0) & 0.794$\pm$0.004(38.0) & 37.38 \\
 & QBC & 0.700$\pm$0.008(37.0) & 0.600$\pm$0.007(37.0) & 0.660$\pm$0.014(38.0) & 0.889$\pm$0.004(35.0) & 0.746$\pm$0.010(32.0) & 0.754$\pm$0.006(36.0) & 0.924$\pm$0.002(35.0) & 0.806$\pm$0.009(37.0) & 35.88\\\hline
\end{tabular}%

}

\end{table}
\end{landscape}

\subsection{Results and Analysis}

\begin{figure*}[h!]

  \centering
  \subfigure[Random \label{fig:MDCR_random}]{\includegraphics[width=0.3\linewidth]{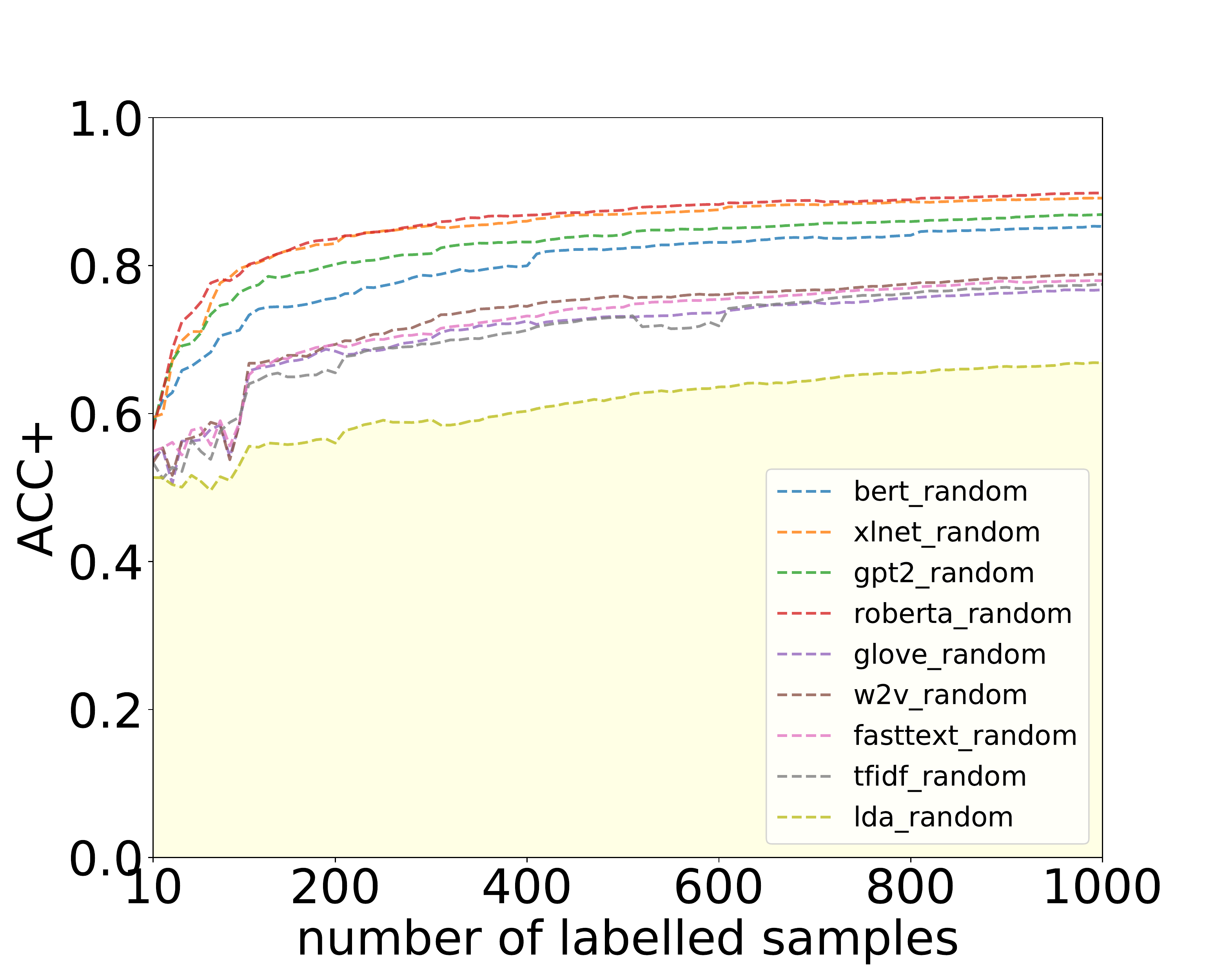}}
  \subfigure[Uncertainty]{\includegraphics[width=0.3\linewidth]{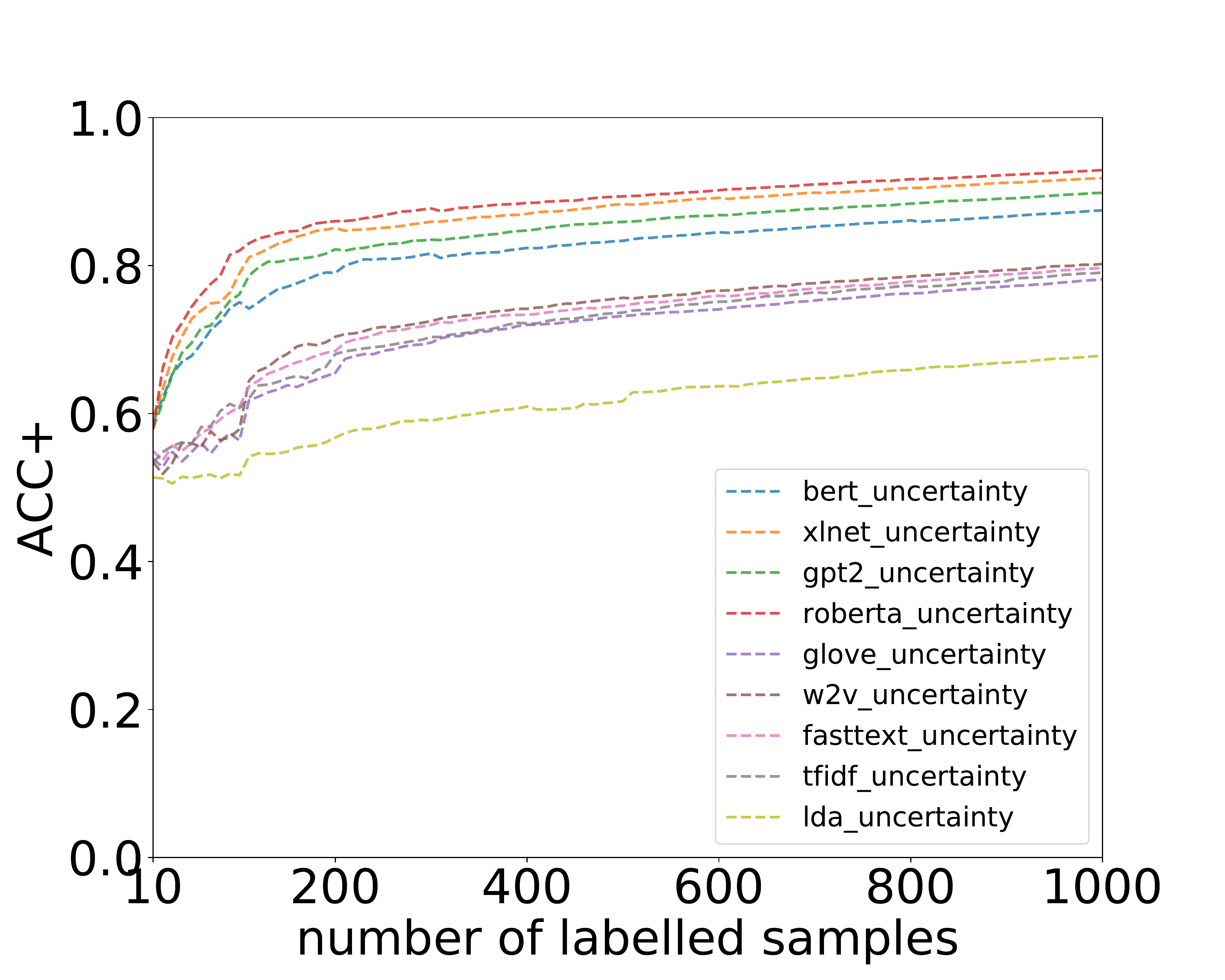}}
\subfigure[Information Density]{\includegraphics[width=0.3\linewidth]{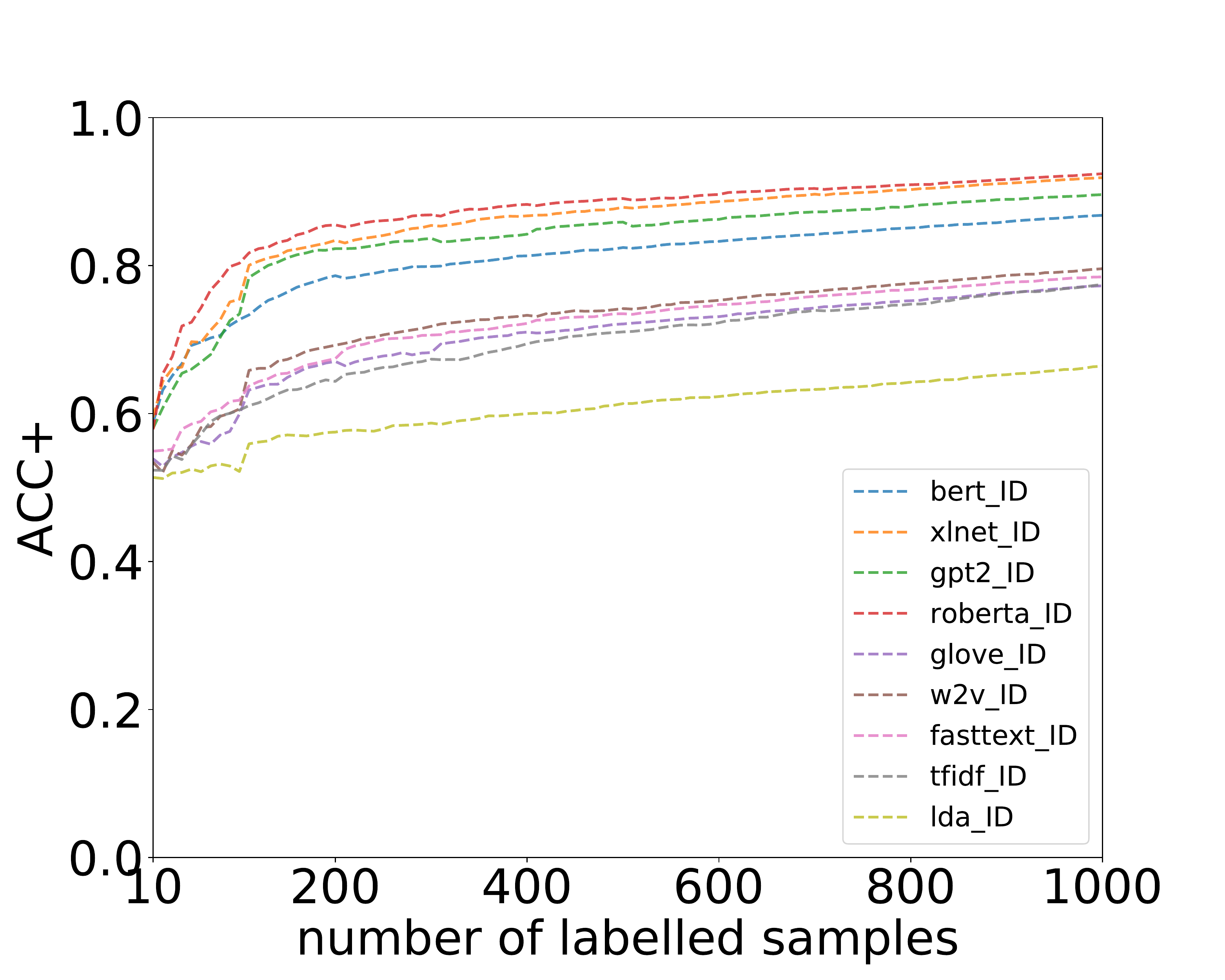}}
\medskip

  \centering
  \subfigure[QBC]{\includegraphics[width=0.3\linewidth]{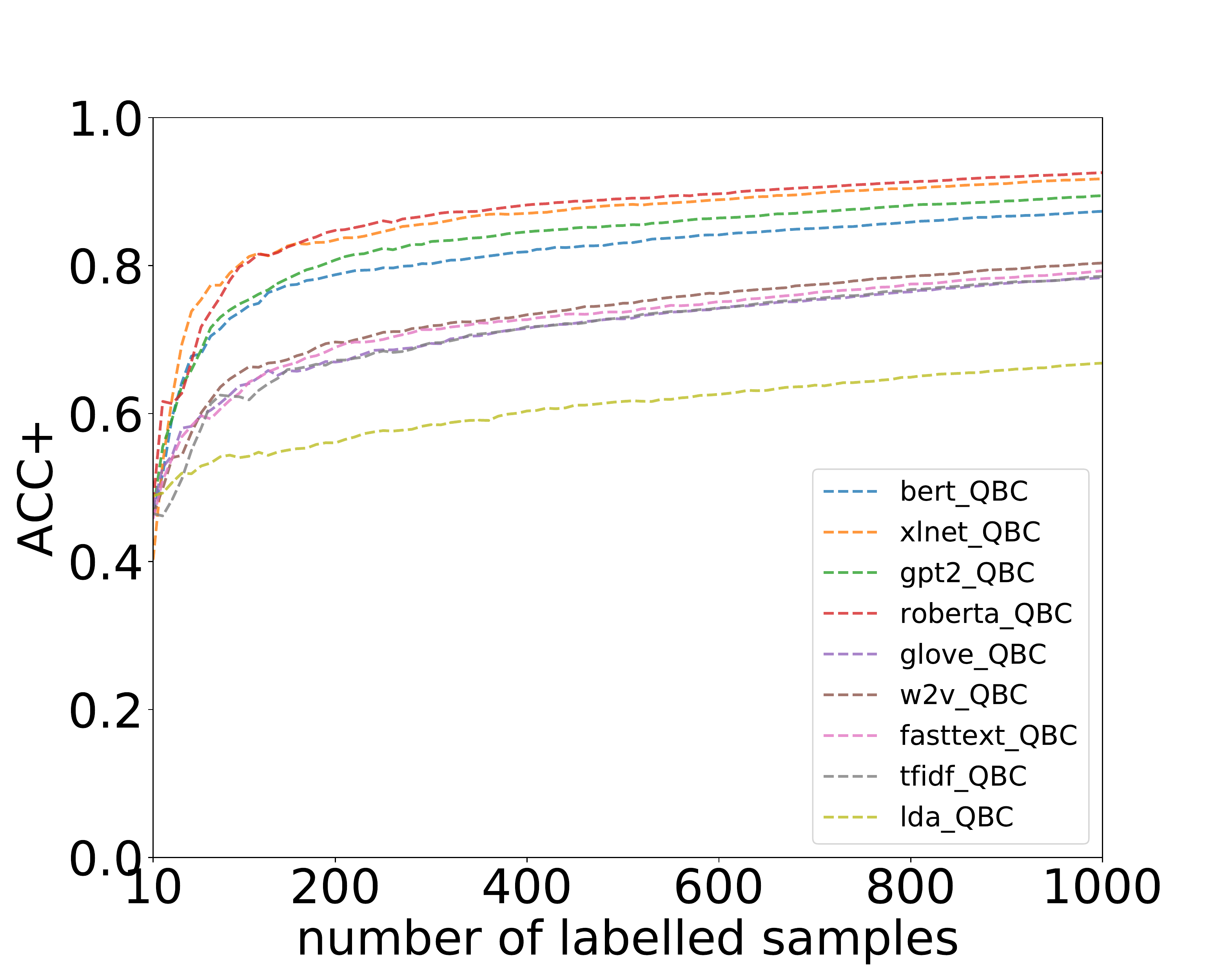}}
  \subfigure[EGAL]{\includegraphics[width=0.3\linewidth]{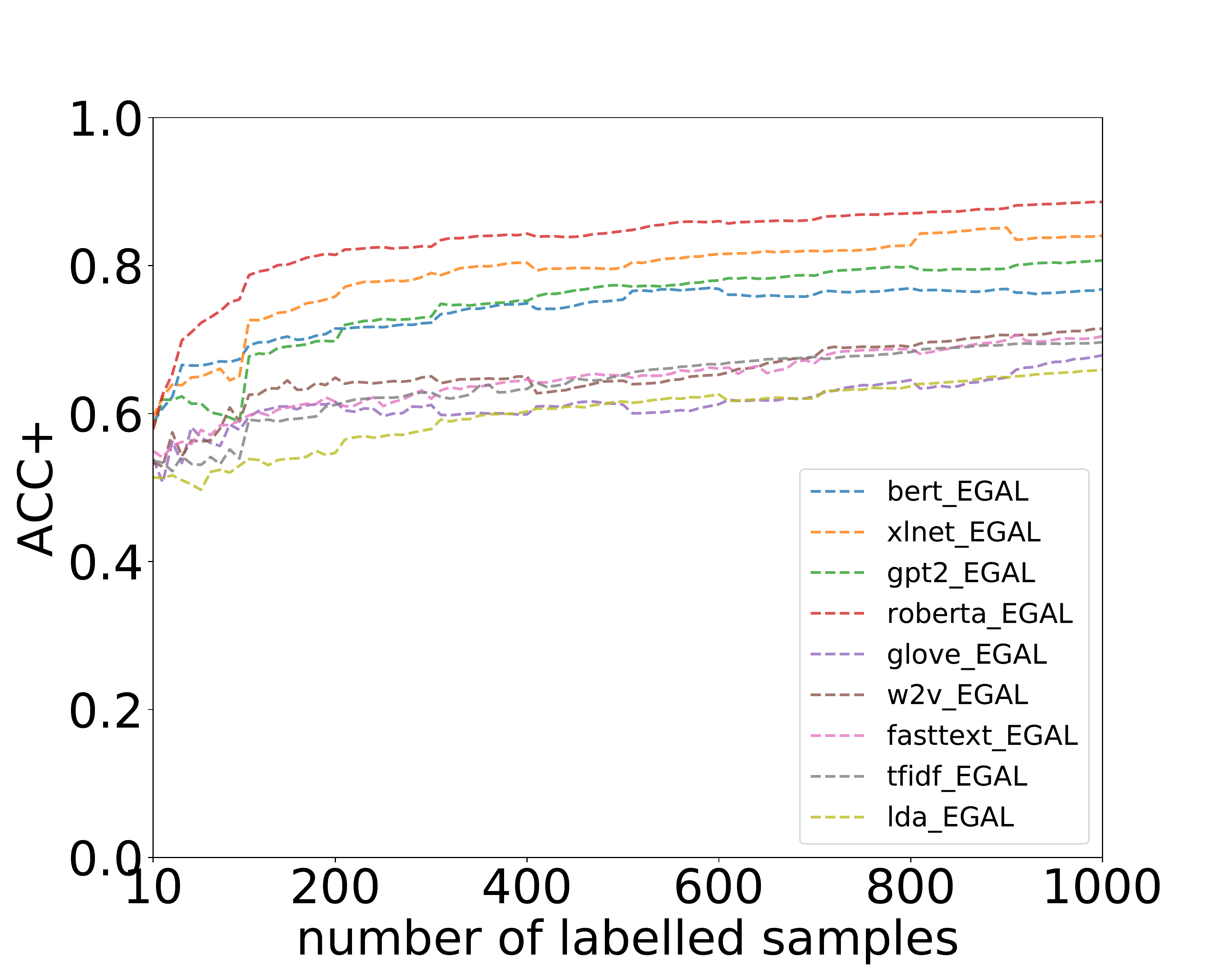}}

  \caption{Results over \emph{Multi-Domain Customer Review (MDCR)} dataset regarding different representation techniques. X-axis represents the number of documents that have been manually annotated and Y-axis denotes accuracy+. Each curve starts with 10 along X-axis.}
  \label{fig:MDCR}

\end{figure*}

\begin{figure*}[h!]
  \centering
  \subfigure[Random]{\includegraphics[width=0.3\linewidth]{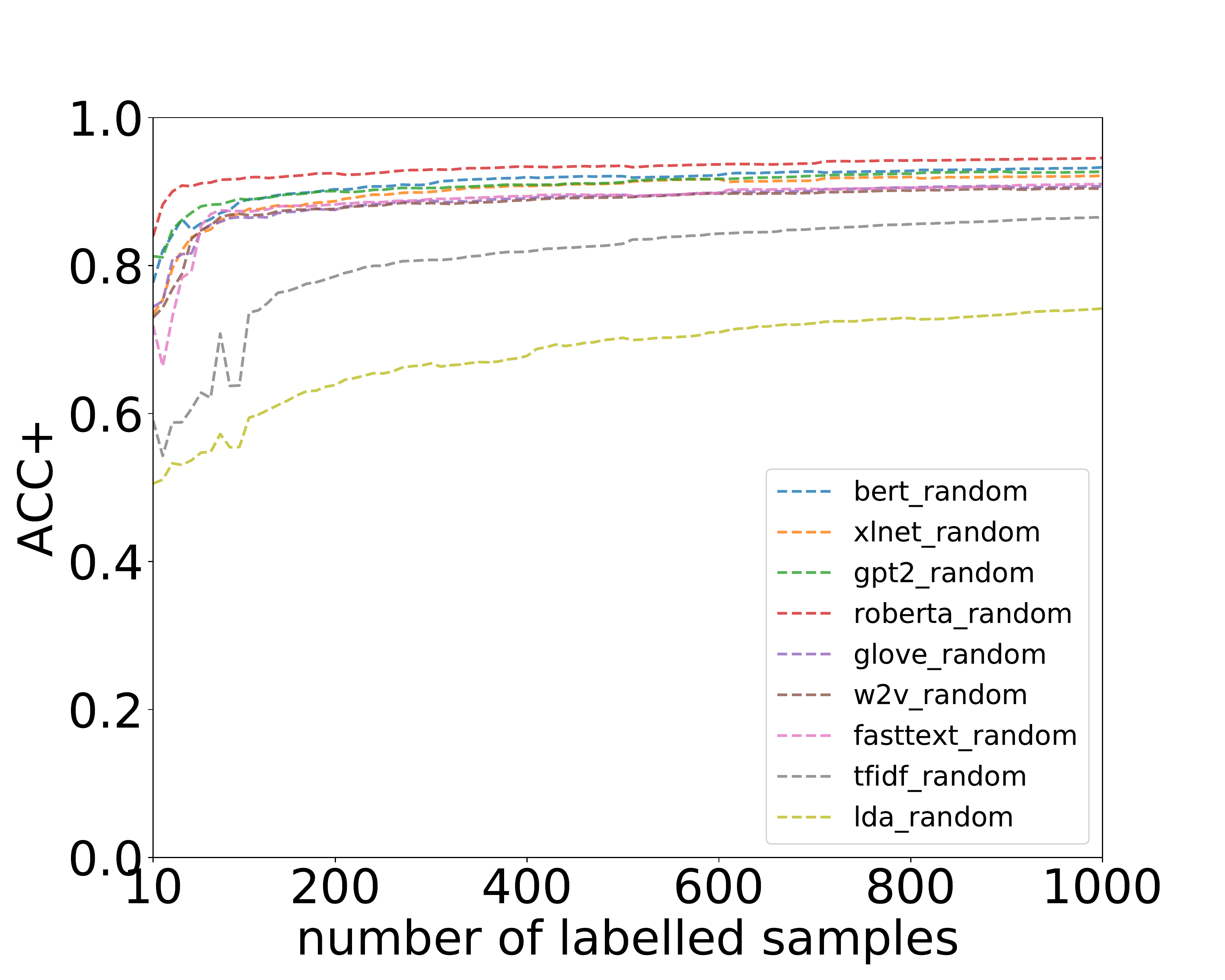}}
  \subfigure[Uncertainty]{\includegraphics[width=0.3\linewidth]{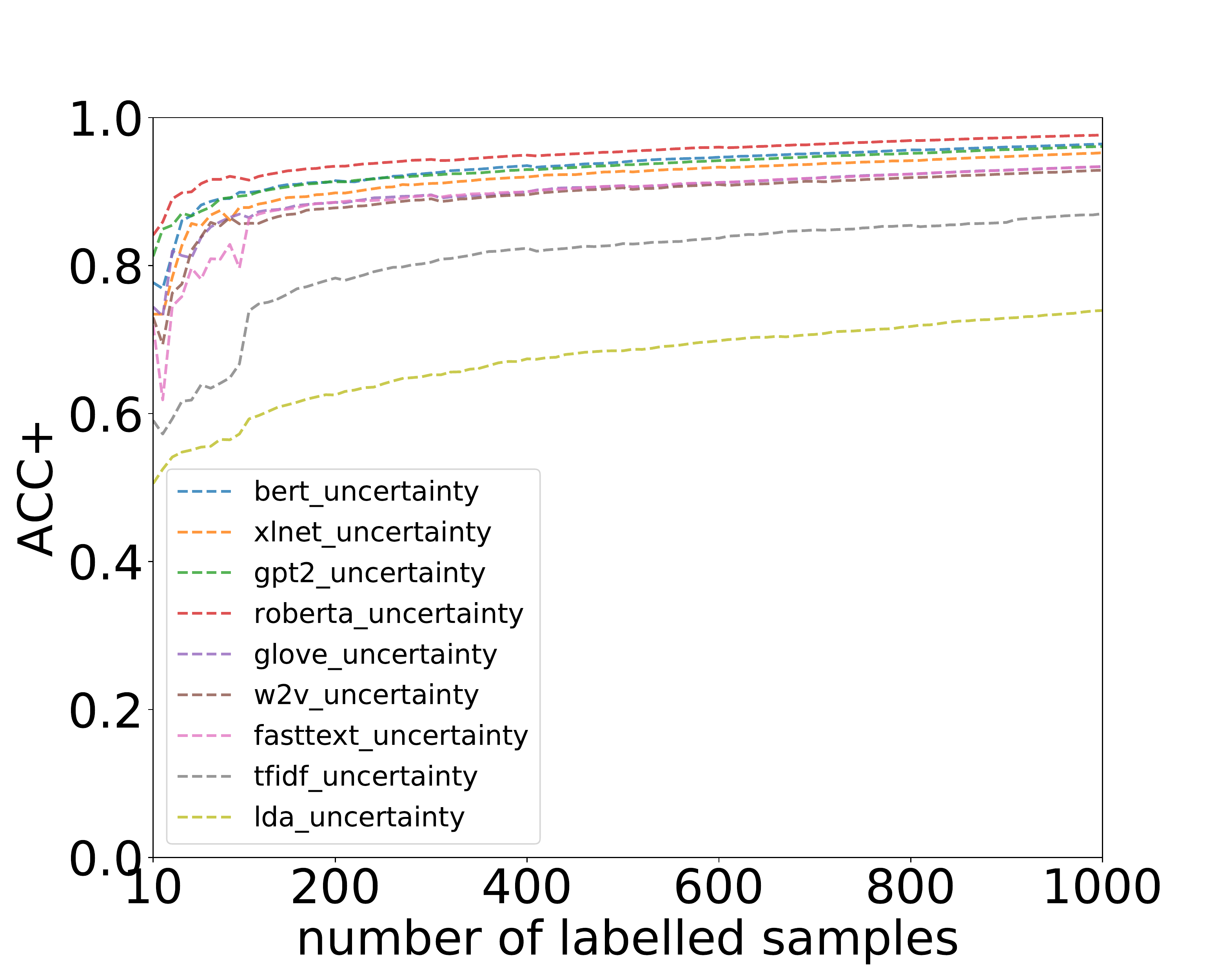}}
\subfigure[Information Density]{\includegraphics[width=0.3\linewidth]{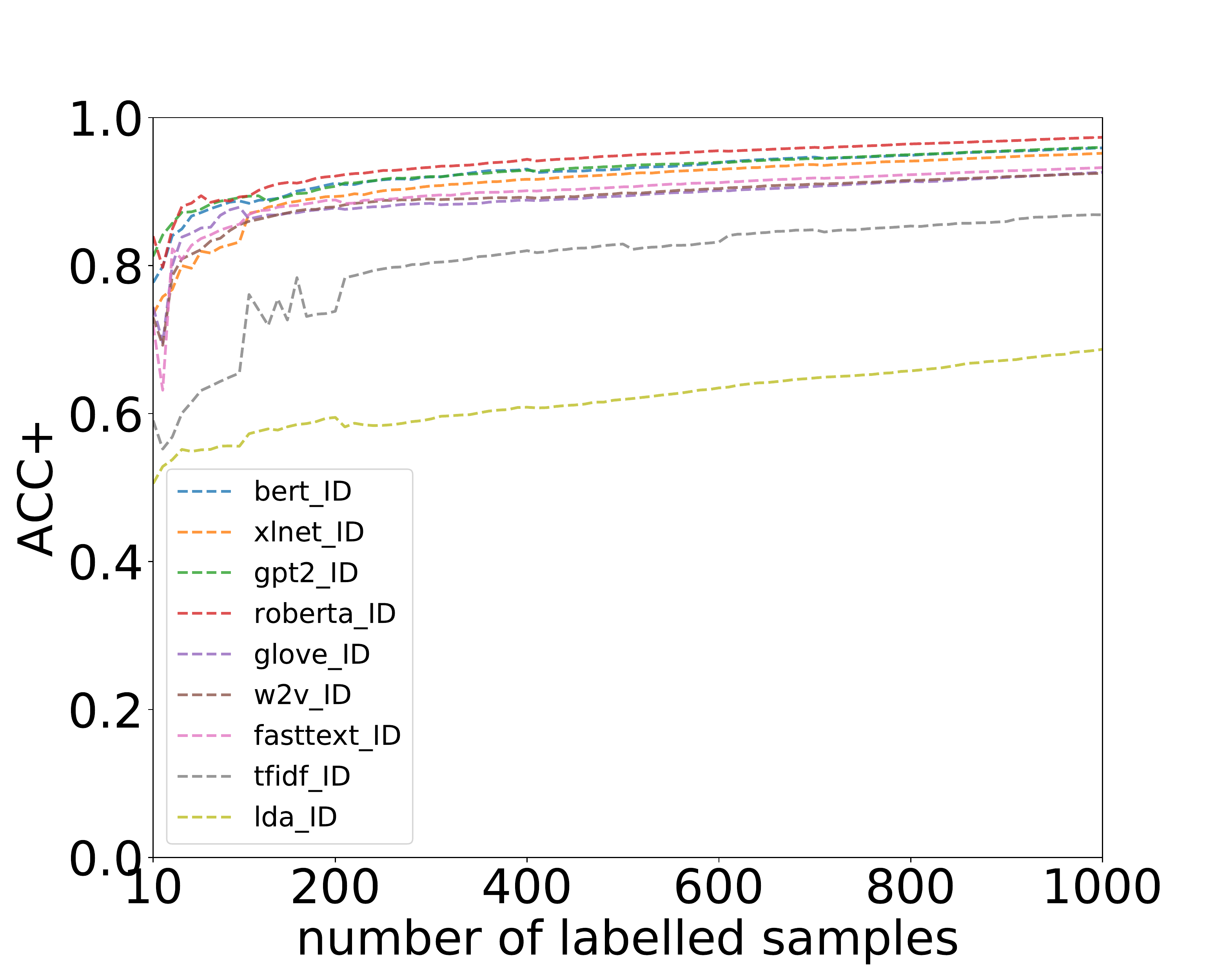}}
\medskip

  \centering
  \subfigure[QBC]{\includegraphics[width=0.3\linewidth]{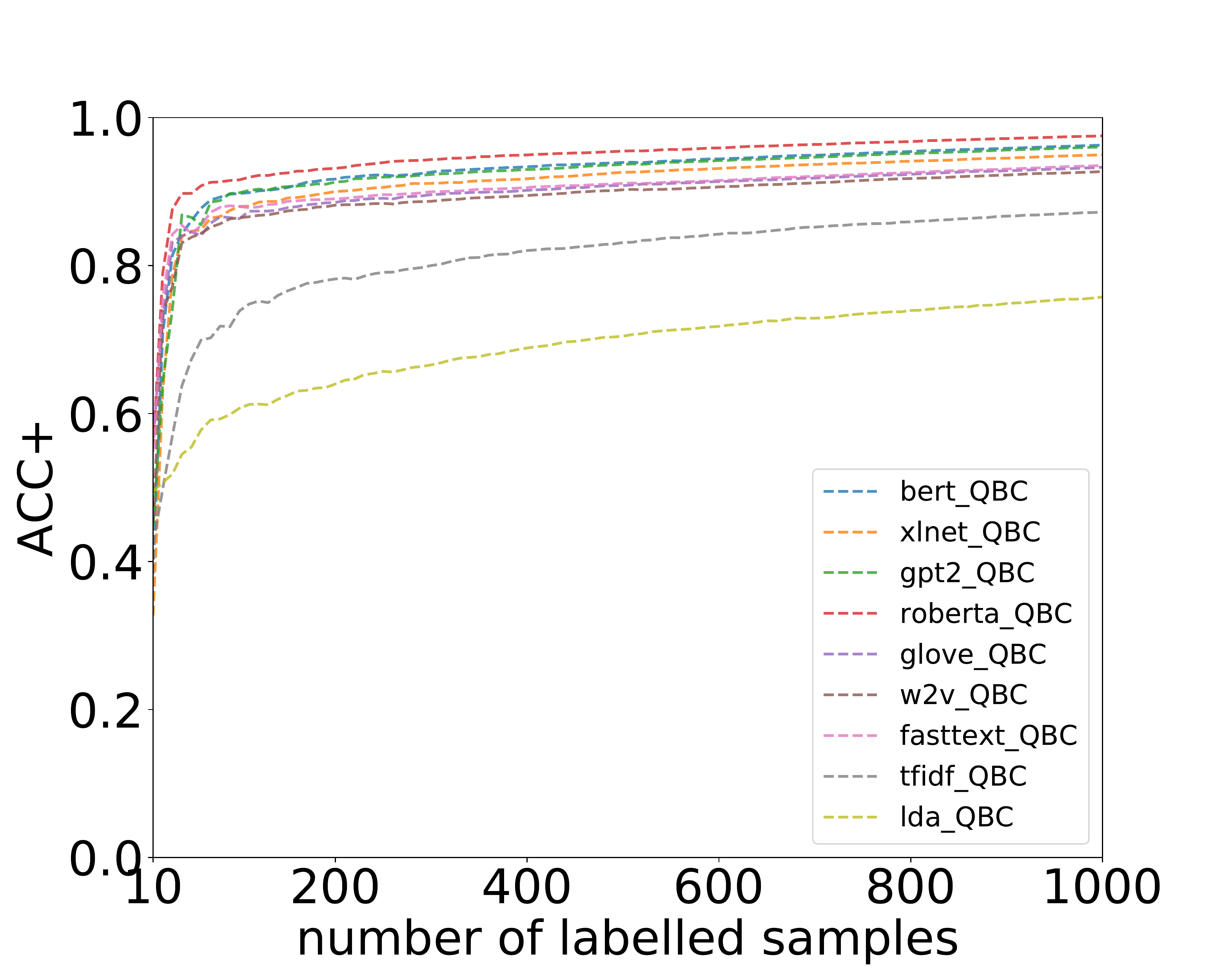}}
  \subfigure[EGAL]{\includegraphics[width=0.3\linewidth]{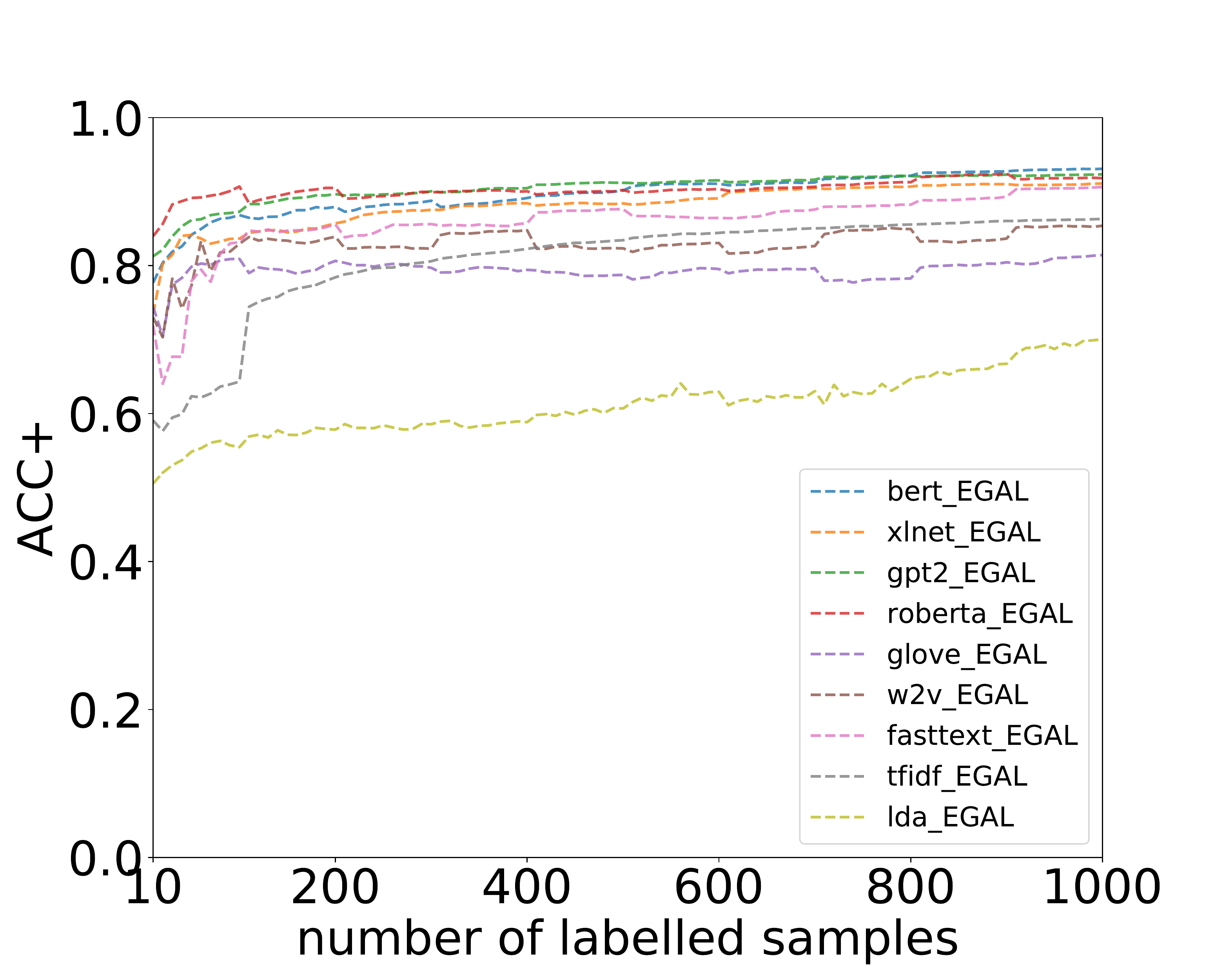}}

  \caption{Results over \emph{Movie Review Subjectivity (MRS)} dataset regarding various representation techniques. X-axis represents the number of documents that have been manually annotated and Y-axis denotes accuracy+. Each curve starts with 10 along X-axis.}
  \label{fig:MRS}

\end{figure*}

To illustrate the performance differences observed between the different representations explored, Figures \ref{fig:MDCR} and \ref{fig:MRS} show the learning curves for each different representation (separated by selection strategy) for the MDCR and MRS datasets respectively. In these plots the horizontal axis denotes the number of instances labelled so far, and the vertical axis denotes the \emph{accuracy+} score achieved. It should be noted that each curve starts with 10 rather than 0 along the horizontal axis, corresponding to the initial seed labelling described earlier. 

Generally speaking, we can observe that better performance is achieved when active learning is used in combination with a text representations based on transformer-based-embeddings rather than the simpler frequency-based  representations (i.e. TF-IDF and LDA), or representations based on word embeddings (i.e. word2vec, Glove, and FastText). More specifically, in Figure \ref{fig:MDCR}, we observe that Roberta consistently outperforms any other representation by a reasonably large margin across all selection strategies. Another interesting observation is that the approaches based on word2vec, Glove and FastText give similar performance, and the approach based on LDA performs worst across all situations. In Figure \ref{fig:MRS}, we see a similar pattern that, in the majority of cases, the performance of the approaches based on transformer-based models surpass the performances achieved using other representations. Besides, Roberta always outperforms other representations except when EGAL is used as the selection strategy. Again, LDA performs poorly when used in combination with all selection strategies.

We collate the AULC results of all methods in Table \ref{tab:AUC_summary}. In this table, each column denotes the performance of different active learning processes on a specific dataset. Different representation and selection strategy combinations are compared, and the best results achieved for each dataset are highlighted. The numbers in brackets stand for the ranking of each method when compared to the performance of the other approaches for a specific dataset. The last column reports the average ranking of each representation-selection-strategy combination, where a smaller number means a higher rank. Table \ref{tab:AUC_summary_small} presents a summarised version where only the selection strategy that gives the best performance of each representation is shown. 

Table \ref{tab:AUC_summary} presents a very clear message that the transformer-based representations perform well across all datasets, which is evidenced by the higher ranks they receive, as compared to TF-IDF, LDA and other word embeddings such as word2vec. Overall, Roberta is the best performing representation with average ranks of 2.12 for Roberta + uncertainty, 4.19 for Roberta + information density, and 6.31 for Roberta + QBC being the highest average ranks overall. This addresses RQ1 defined in /Section \ref{sec:intro} and clearly shows that transformer-based representations should be used in active learning systems for text classification.

\section{Which is the best BERT?} \label{sec:experiment_2}

This section describes an experiment designed to answer two questions. First, are the representations generated using the lightweight versions of BERT and BERT-like models as effective as the those generated using the regular models (RQ2 from Section \ref{sec:intro}). Second, when using a representation from a transfomer-based model in active learning for text classification are representations based on the ``[CLS]'' token more effective than those generated by  aggregating word representations?  The setup of the experimental framework, the performance measures used, and the datasets used are the same as those mentioned in Sections \ref{sec:active_setting}, \ref{sec:metric} and \ref{sec:dataset}. The rest of the subsections describes the set up of the language models, the results and corresponding analysis.

\subsection{Configurations of Text Representation Techniques}

We adopted the original BERT (``bert-base-uncased'') as a baseline in this experiment. This is compared to Roberta (``roberta-base'') and two commonly used lighter version of BERT: ALbert (``albert-base-v2'') and Distilbert (``distilbert-base-uncased''). The codes beside the algorithms are the specific models used in the experiments which can be found on Github. \footnote{https://github.com/huggingface/transformers}
For each model, texts are represented by both ``[CLS]'' token and averaged embeddings. In other words, there are 8 different representations being compared in these experiments. For the averaged representations, if the text sequence is longer than 512 tokens, the same aggregated method is applied as described in Section \ref{sec:text_configurations}. Similarly, for ``[CLS]'' representation, the document is encoded by the mean of the ``[CLS]'' token embeddings for each fraction.

\subsection{Results and Analysis}

\begin{figure*}[h!]
  \centering
  \subfigure[Random ]{\includegraphics[width=0.27\linewidth]{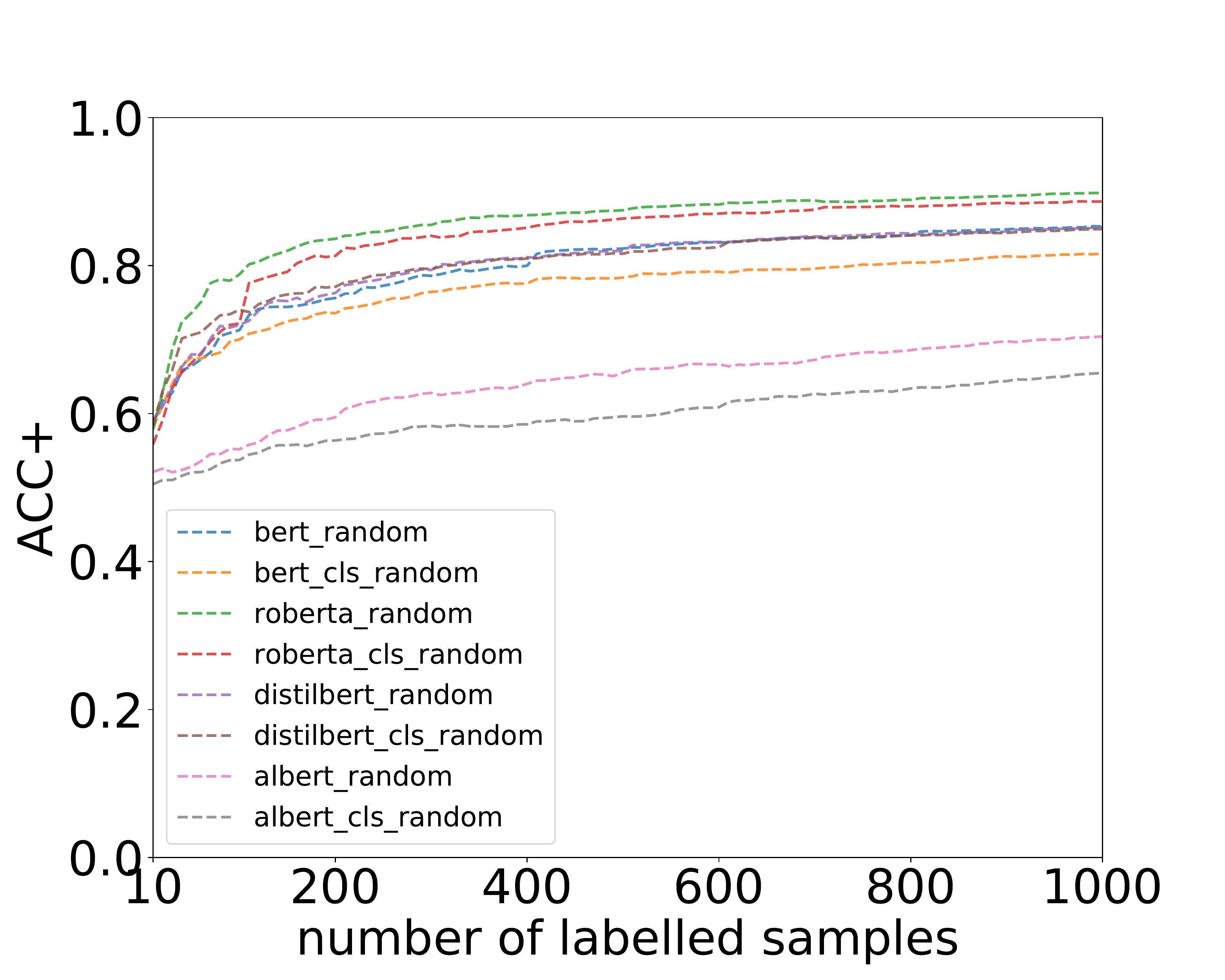}}
  \subfigure[Uncertainty]{\includegraphics[width=0.27\linewidth]{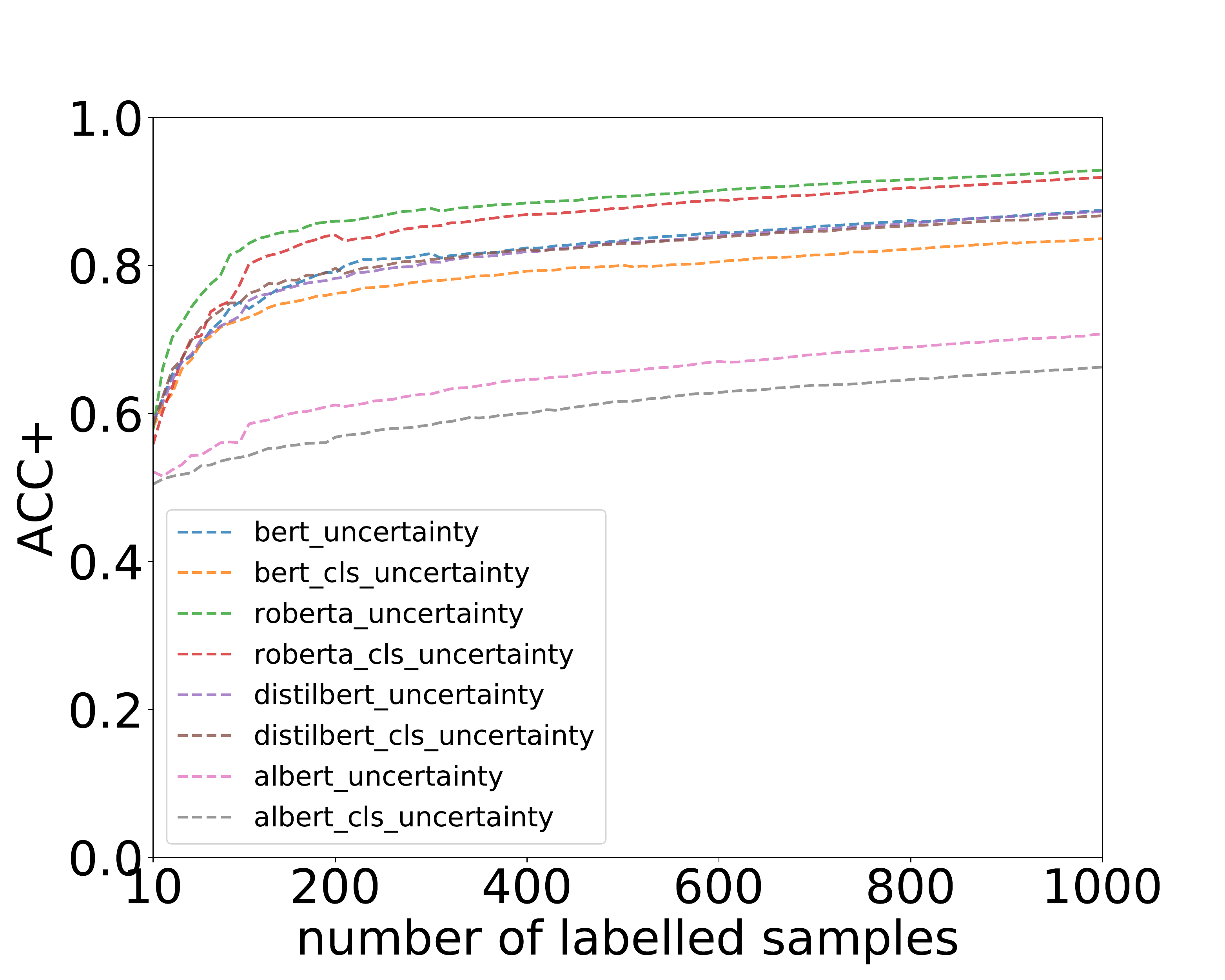}}
\subfigure[Information Density]{\includegraphics[width=0.27\linewidth]{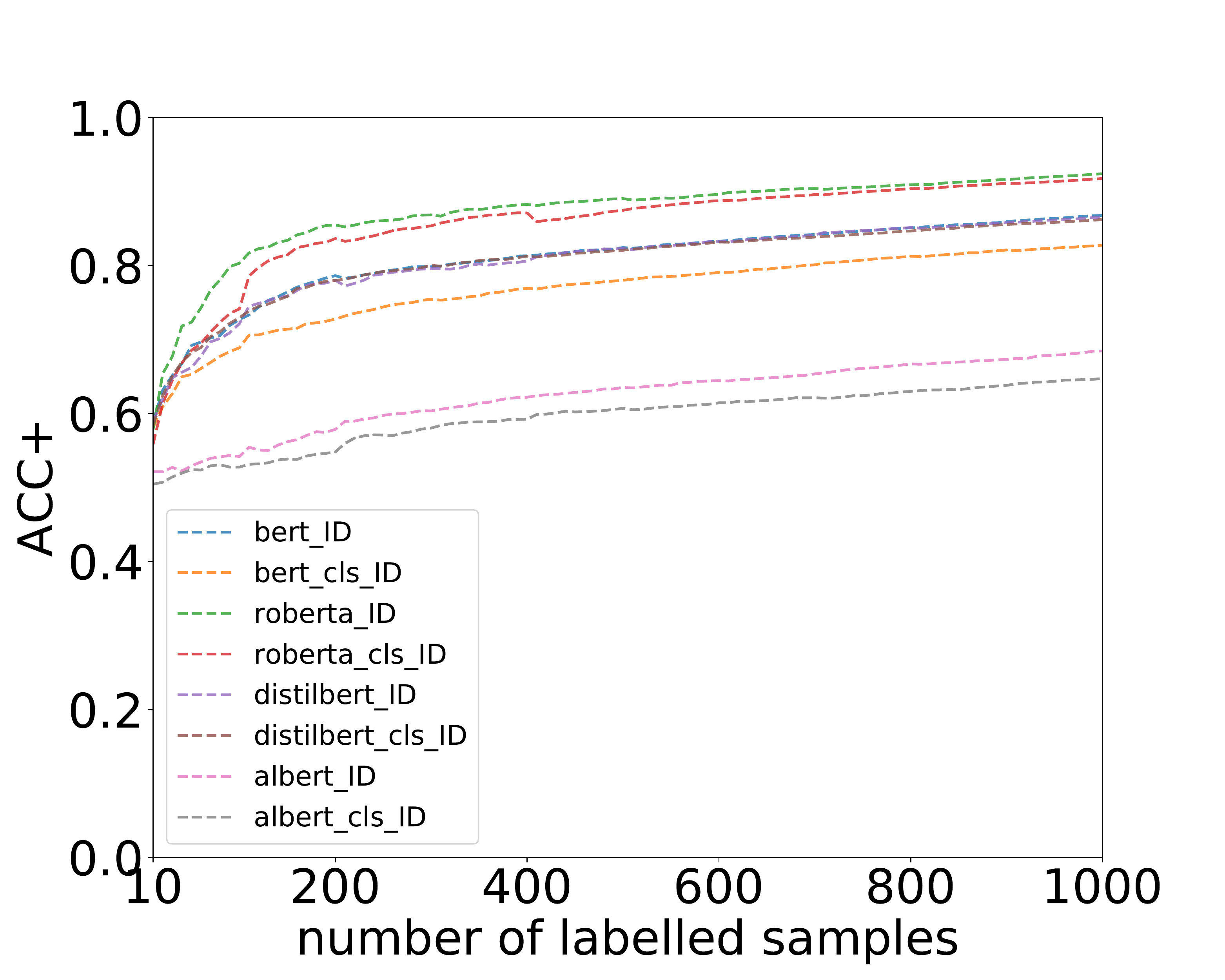}}
\medskip

  \centering
  \subfigure[QBC]{\includegraphics[width=0.27\linewidth]{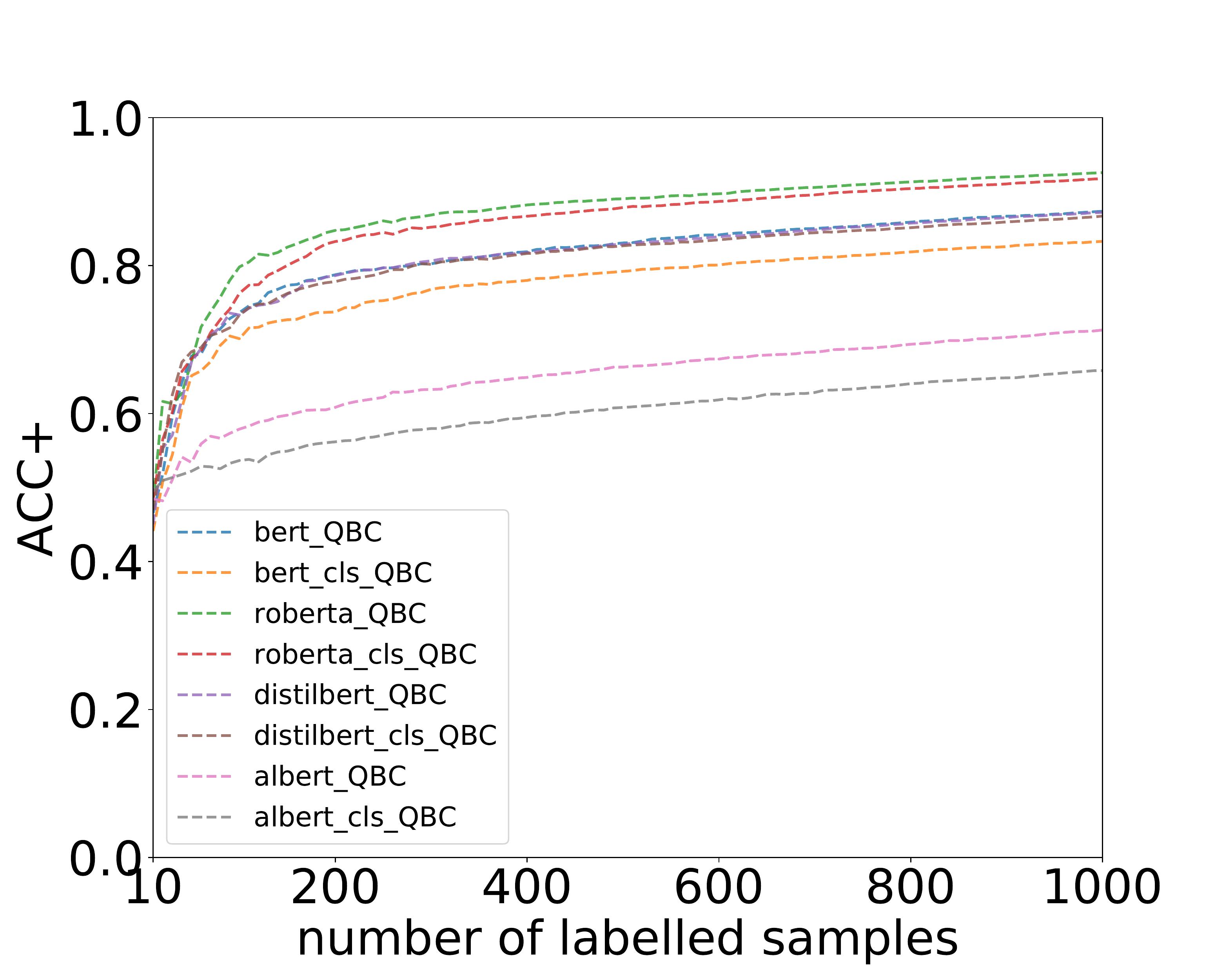}}
  \subfigure[EGAL]{\includegraphics[width=0.27\linewidth]{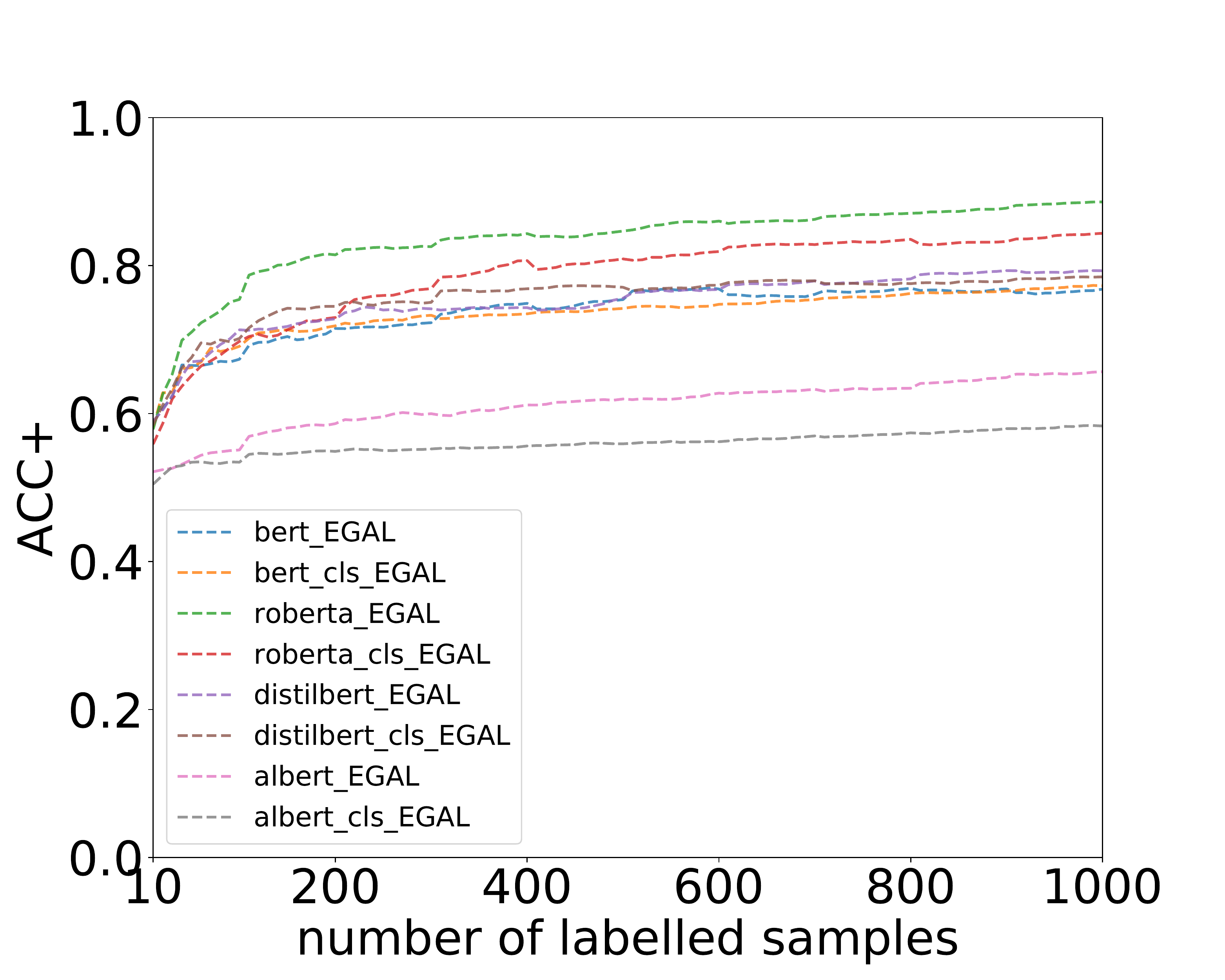}}

  \caption{Results over \emph{Multi-Domain Customer Review (MDCR)} dataset regarding different variants of BERT. X-axis represents the number of documents that have been manually annotated and Y-axis denotes accuracy+. Each curve starts with 10 along X-axis.}
  \label{fig:MDCR_cls}
  
\end{figure*}

\begin{figure*}[h!]
  \centering
  \subfigure[Random]{\includegraphics[width=0.3\linewidth]{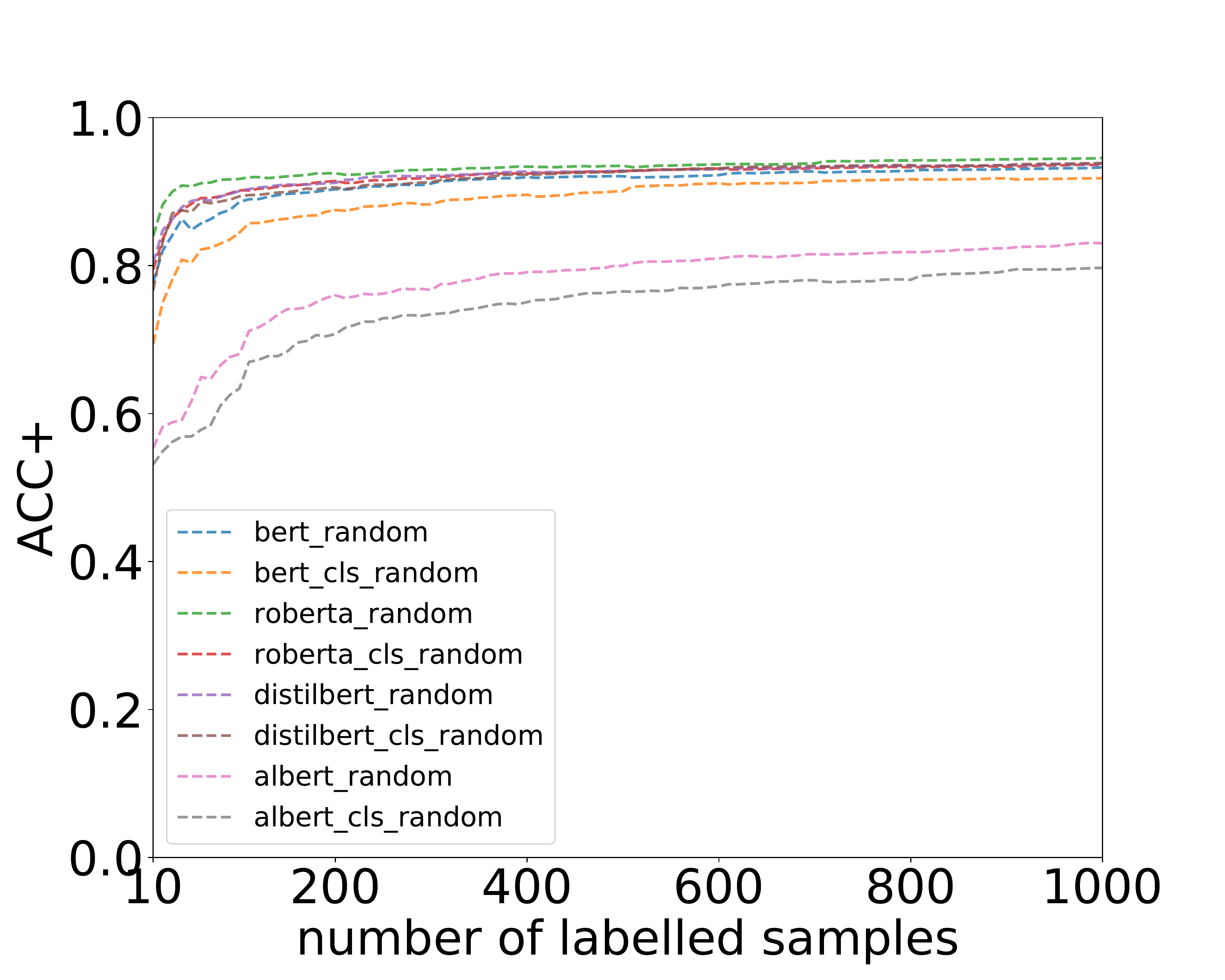}}
  \subfigure[Uncertainty]{\includegraphics[width=0.3\linewidth]{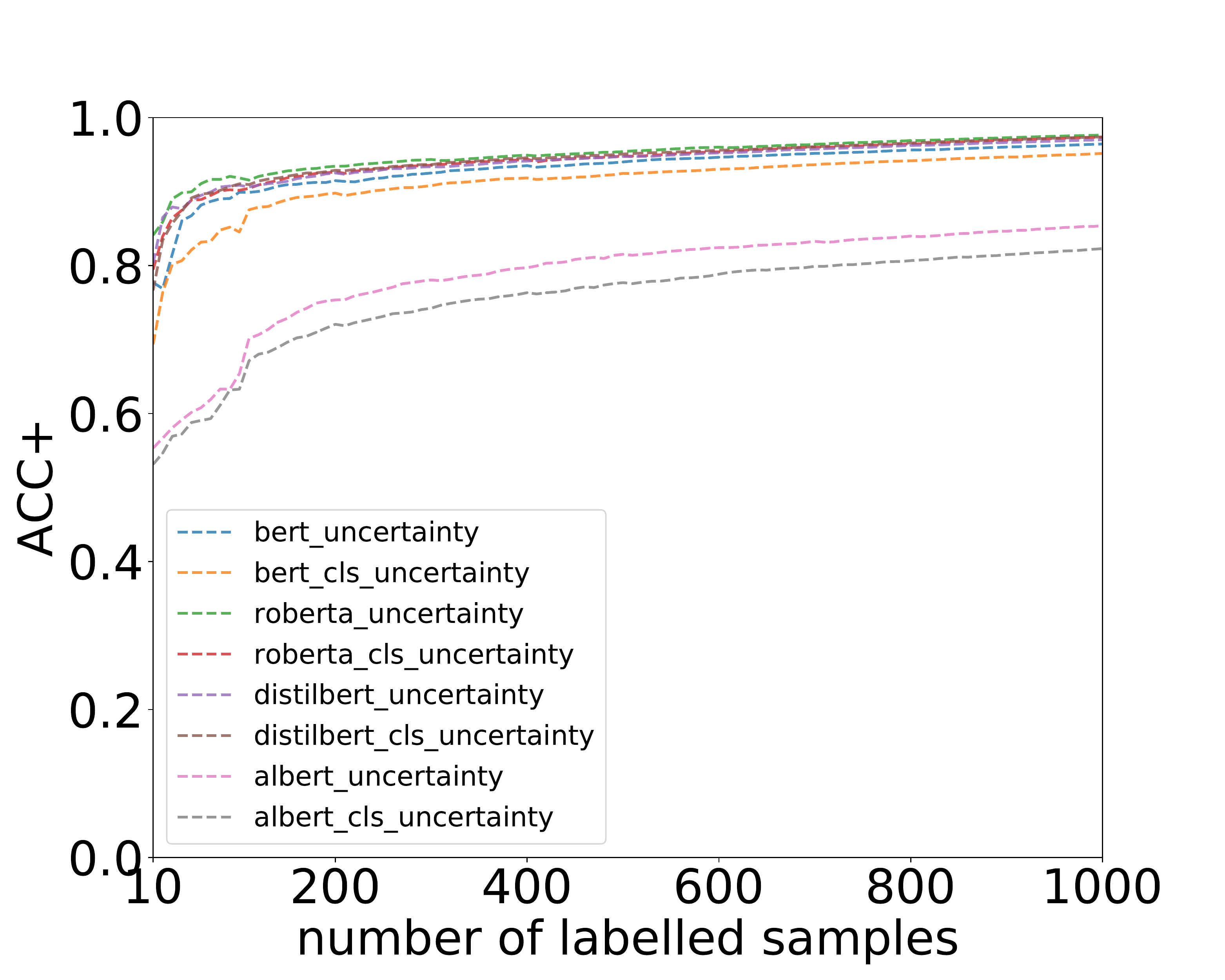}}
\subfigure[Information Density]{\includegraphics[width=0.3\linewidth]{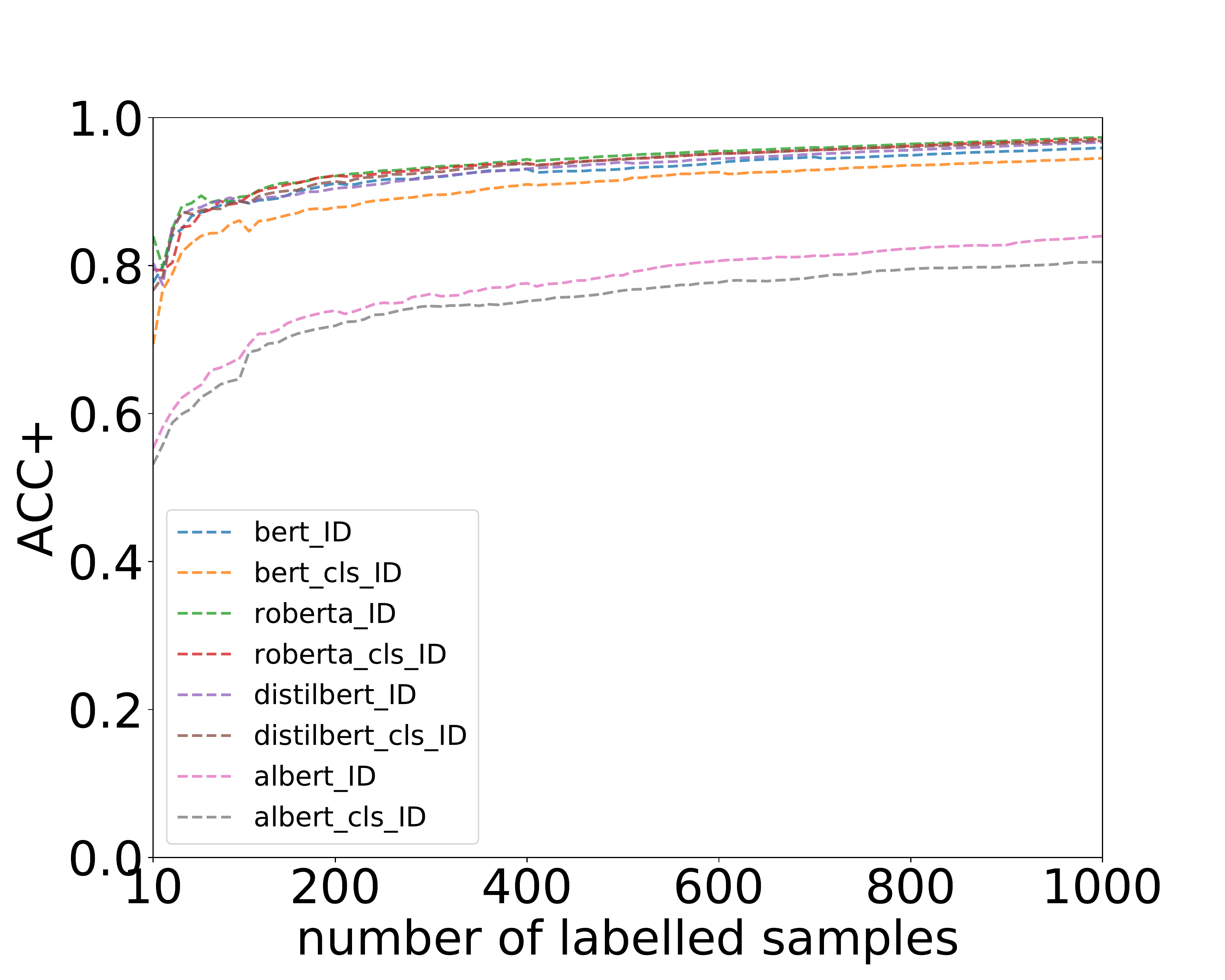}}
\medskip

  \centering
  \subfigure[QBC]{\includegraphics[width=0.3\linewidth]{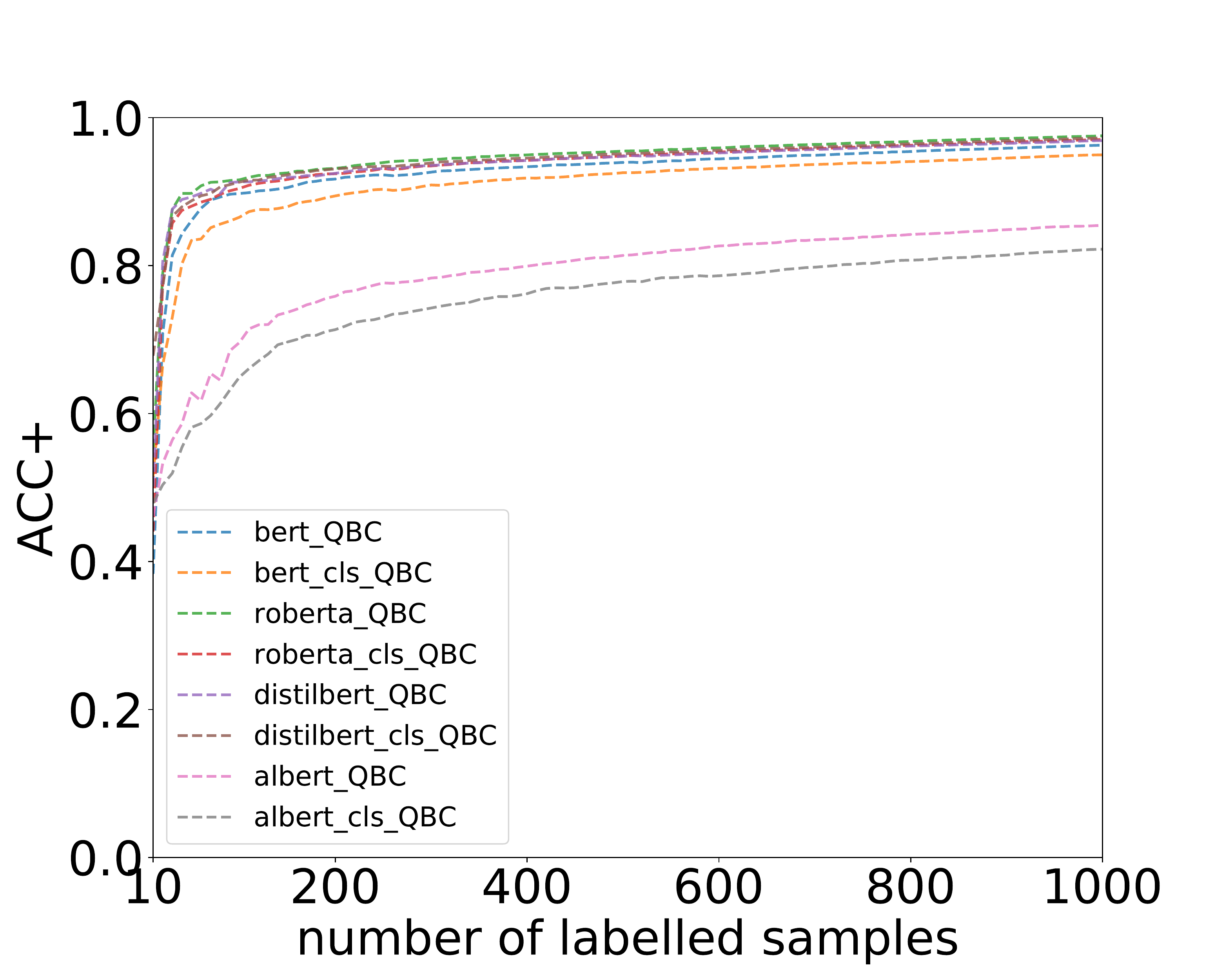}}
  \subfigure[EGAL]{\includegraphics[width=0.3\linewidth]{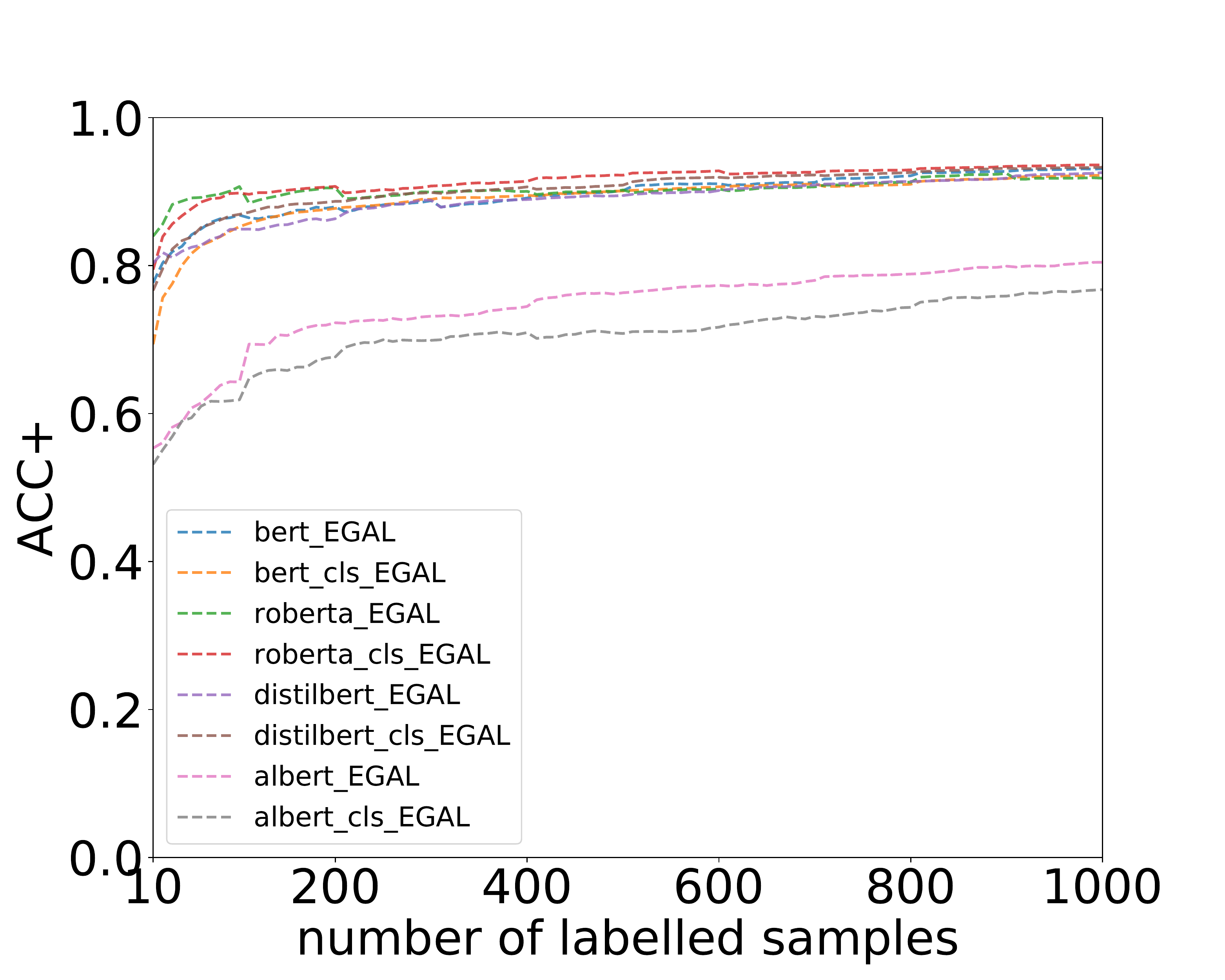}}

  \caption{Results over \emph{Movie Review Subjectivity (MRS)} dataset regarding different variants of BERT. X-axis represents the number of documents that have been manually annotated and Y-axis denotes accuracy+. Each curve starts with 10 along X-axis.}
  \label{fig:MRS_cls}

\end{figure*}

To demonstrate the effectiveness of variants of BERT explored and the impacts of averaged representation and ``[CLS]'' representation, Figures \ref{fig:MDCR_cls} and \ref{fig:MRS_cls} show the learning curves for each different representation (separated by selection strategy) for the MDCR and MRS datasets respectively.
One obvious observation is that the Roberta model is still the best pre-trained language model across all situations. Another interesting fact is that within the same language model, the averaged representation is always better than the ``[CLS]'' token representation. This matches the finding by Devlin et al. (\citeyear{devlin2018bert}) that the ``[CLS]'' representation could manifest its performance only if the model is fine-tuned. 

More specifically, the original BERT and the smaller DistilBert have a similar performance and surprisingly, Albert fails to demonstrate the competitive performance as compared to other models. These observations can also be seen in the AULC score summarized in Table \ref{tab:AUC_CLS_summary}. The ranks of ``[CLS]'' representations are always lower than that of their averaged counterparts and the ranks of representations based on BERT are close to those based on DistilBert counterparts. Table \ref{tab:gpu_inference_time} reports the number of parameters as well as the inference time cost of various transformer-based models using GPU. DistilBert is always the fastest, while GPT-2 and XLNet are the two most time-consuming models. This is not surprising due to the complicated design of GPT-2 and XLNet. BERT and Roberta have a similar inference time which is expected as they share the same model architecture. Though the number of parameters in Albert is much less than that in Roberta and original BERT, the inference speed of Albert is nearly the same as that of Roberta and BERT. We suppose that this is due to the matrix multiplication process involved in Albert. Considering both the performance in accuracy and inference speed, we recommend using Roberta + averaged representation + uncertainty as to the default active learning setting, however, if the user has a problem with GPU memory or pursue a fast inference speed, DistilBert could be an effective alternative. These findings address RQ2 and RQ3 defined in Section \ref{sec:intro}.



\section{Adaptive Tuning Active Learning} \label{sec:experiment_3}

One thing that distinguishes the representations based on transformer-based models from those based on simpler word embeddings (e.g. word2vec) is that the transformer-based models can use fine-tuning to take advantage of label information that arises during active learning. Although a relatively small number of labels are generated during active learning it may be possible to use them to learn more relevant embeddings that will improve the overall active learning process (RQ4 from Section \ref{sec:intro}). To this end we propose \emph{Adaptive Tuning Active Learning} (ATAL), an  algorithm where the pre-trained language model is adaptively improved via fine-tuning with label information as it becomes available during active learning. To the best of our knowledge, this is the first approach to exploiting the label information for building better representations during the active learning process. The following subsections describe the methodology, the experimental configurations, results and analysis.

\subsection{Method}

Algorithm \ref{alg:ATAL} describes ATAL in detail. ATAL is built on the standard pool-based active learning procedure and the algorithm is similar to Algorithm \ref{alg:AL}. The main difference is that on Lines 13 to 18 in Algorithm \ref{alg:ATAL} we fine-tune the transformer-based model every 20 iterations (i.e. for every 200 newly labelled instances, assuming a batch size of 10) with currently labelled instances. 
As the amount of labelled data available for fine tuning is small we do not have the luxury of a hold out validation set to use to implement early stopping during model fine tuning. Instead after training for 15 epochs we roll back to the model with the lowest loss based on the training dataset. This is not ideal and runs the risk of overfitting, however, our experimental results show that it is effective.  
Another key characteristic of ATAL is that the document representations are changed after each fine-tuning, while in Algorithm \ref{alg:AL}, the document vectors are fixed across all loops. 

We evaluate the performance of ATAL through an evaluation experiment using the same  datasets and using the same performance measures as those described in Sections  \ref{sec:dataset} and \ref{sec:metric}. Again, the ATAL process is repeated 10 times using different random seeds and the performance measures reported are averaged across these repetitions.

\begin{algorithm}[h!]
\DontPrintSemicolon
\KwIn{$T$, set of all corpus \\
    $P$, index of all ground truth positive documents \\
    $N$, index of all ground truth negative documents \\
    $LM$, pre-trained language model}
\KwOut{$S$, set of accuracy+ scores of each loop\\
$L$, set of documents pseudo labelled by the oracle \\
 $R$, set of documents labelled by the classifier\\}
  
\SetKwBlock{Init}{Initialization}{}{}
\Init{  
        // \hspace{0.15cm}Infer document vectors\\
        $E\leftarrow Inference(T,LM)$;\\
        // \hspace{0.15cm}Random sampling 5 neg and 5 pos\\
        $L_{t}\leftarrow Random(T,P,5)\cup Random(T,N,5)$;\\
        $L\leftarrow Inference(L_{t},LM)$;\\
        
        $\neg L\leftarrow E\setminus L$;\\
        $\neg L_{t}\leftarrow T\setminus L_{t}$;\\
        $R\leftarrow \emptyset$;  \\
        $loop\leftarrow 0$;\\
        $Params\leftarrow \emptyset$;\\
        $S\leftarrow \emptyset$;
}
\SetKwRepeat{Do}{do}{while}
\While{$\left |L  \right | \le 1000$}{

  \uIf{$loop\mod20==0$ \vspace{1pt}\textbf{and} \vspace{1pt}$loop>0$}{
  //\hspace{0.15cm} Fine-tuning model\\
 $LM\leftarrow Fine\_tune(L_{t},LM,P,N)$;\\
 //\hspace{0.15cm} Re-infer all document vectors\\
 $L\leftarrow Inference(L_{t},LM)$;\\
 $\neg L\leftarrow Inference(\neg L_{t},LM)$;
 }
 
 $CL,Params\leftarrow Train(loop,L,P,N,Params)$; \\
 $X\leftarrow Query(CL,\neg L)$;\\
 $L,\neg L,R\leftarrow Assign(CL,L,X)$; \\
 //\hspace{0.15cm} Compute accuracy+ score\\
 $S\leftarrow S\cup Eval(L,R,P,N)$;\\
 $loop\leftarrow loop + 1$;
}
\SetKwBlock{Begin}{Function}{end function}
\Begin($\text{Fine\_tune} {(} L_{t},LM,P,N {)}$)
{
  $LM\_candidate\leftarrow LM$;\\
  // \hspace{0.15cm} Validate the model with labelled examples\\
  $acc\_candidate\leftarrow Validate(L_{t},LM,P,N)$;\\
  $epoch\leftarrow 0$;\\
    \While{$epoch<15$}
    {
     //\hspace{0.15cm} Update model weights\\
      $LM\leftarrow Forward\_backprop(L_{t},LM,P,N)$;\\
      $acc\leftarrow Validate(L_{t},LM,P,N)$;\\
      \uIf{$acc\_candidate<acc$}{
        $acc\_candidate\leftarrow acc$;\\
        $LM\_candidate\leftarrow LM$;
      }
      $epoch\leftarrow epoch+1$;
      
    }
    
  \Return{$LM\_candidate$}
}
\caption{Pseudo Code for ATAL.}\label{alg:ATAL}
\end{algorithm}

\subsection{Configurations}

Given the fact that Roberta + averaged representation + uncertainty has shown to be very effective in the previous experiments, we chose this as the baseline in this experiment and built ATAL using Roberta. During the fine-tuning, we set the learning rate to $1e-5$, use 15 epochs, set the batch size to 4, and use the Adam optimizer \citep{kingma2014adam} with epsilon equals to $1e-8$. The values of other hyper-parameters follow the default settings in the Roberta model.

As well as reporting the results  of the ATAL model, in this experiment we also report the performance of a Roberta model fine-tuned with the fully labelled dataset, referred to as ``roberta\_tuned''. It would not be possible to use this approach in practice as it would not have access to the fully labelled datasets, but it does provide an interesting upper bound on the possible performance of ATAL.


\subsection{Results and Analysis}

\begin{figure*}[h!]
  \centering
  \subfigure[Movie Review]{\includegraphics[width=0.24\linewidth]{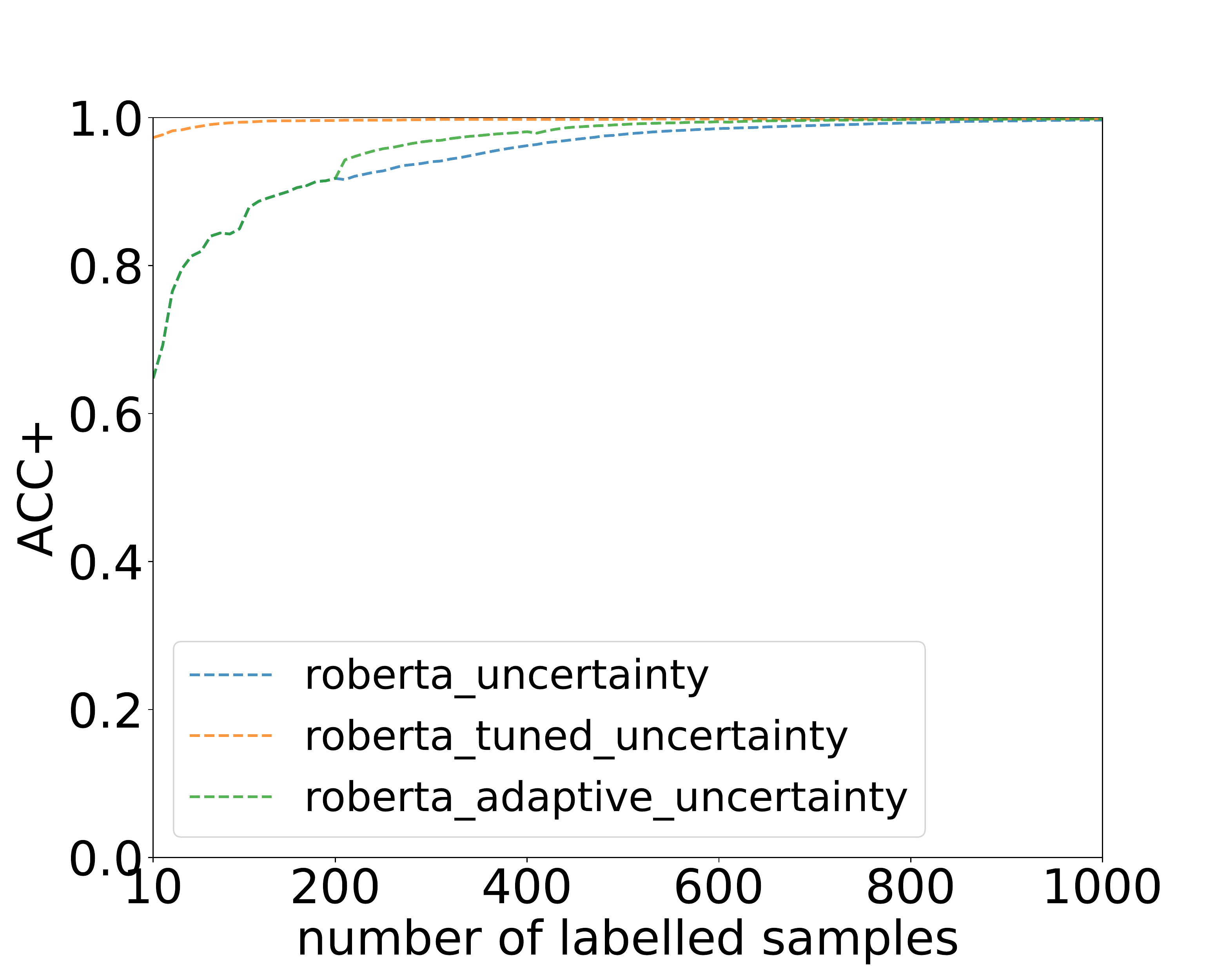}}
  \subfigure[Multi-domain Customer Review]{\includegraphics[width=0.24\linewidth]{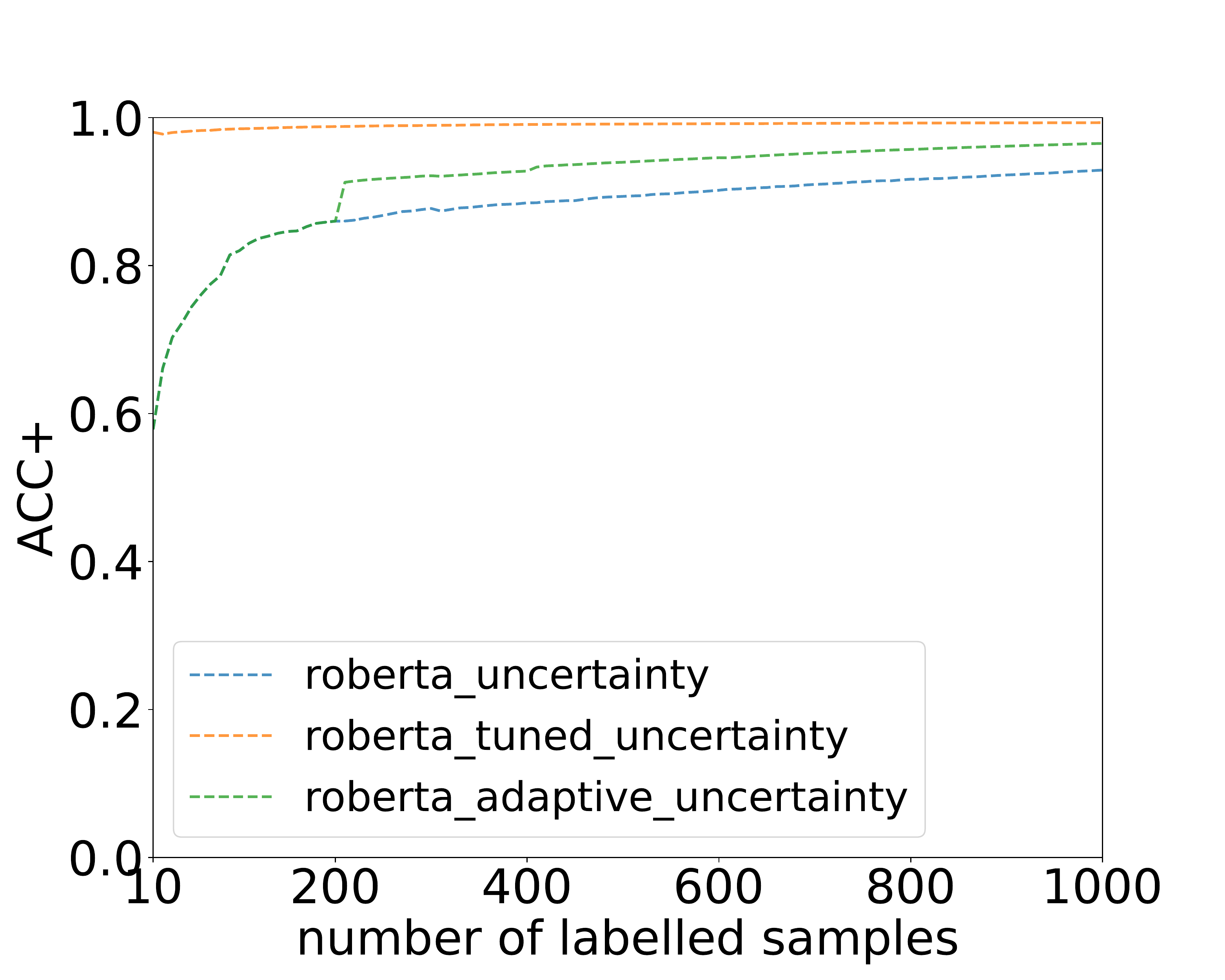}}
\subfigure[Blog Author Gender]{\includegraphics[width=0.24\linewidth]{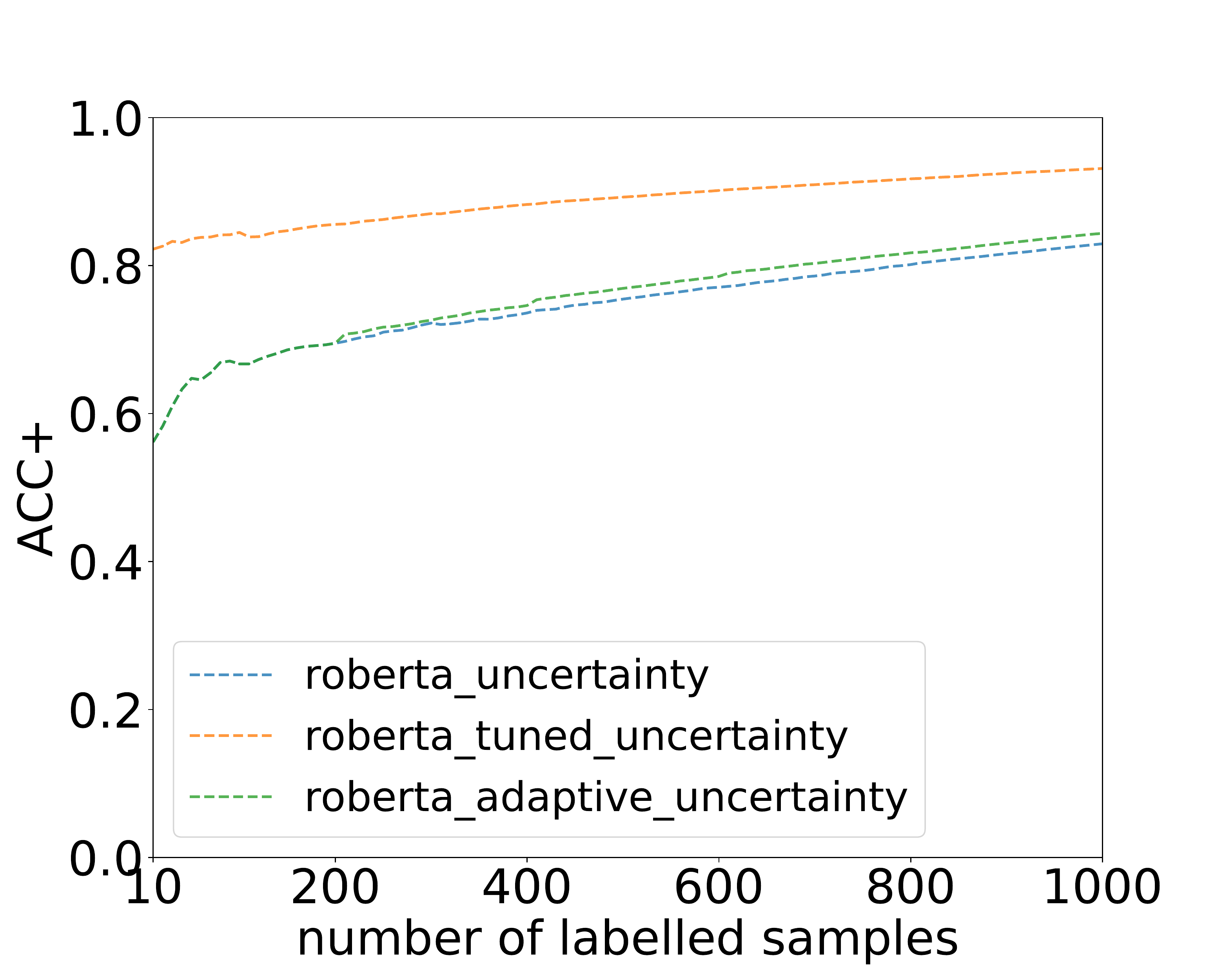}}
\subfigure[Guardian 2013]{\includegraphics[width=0.24\linewidth]{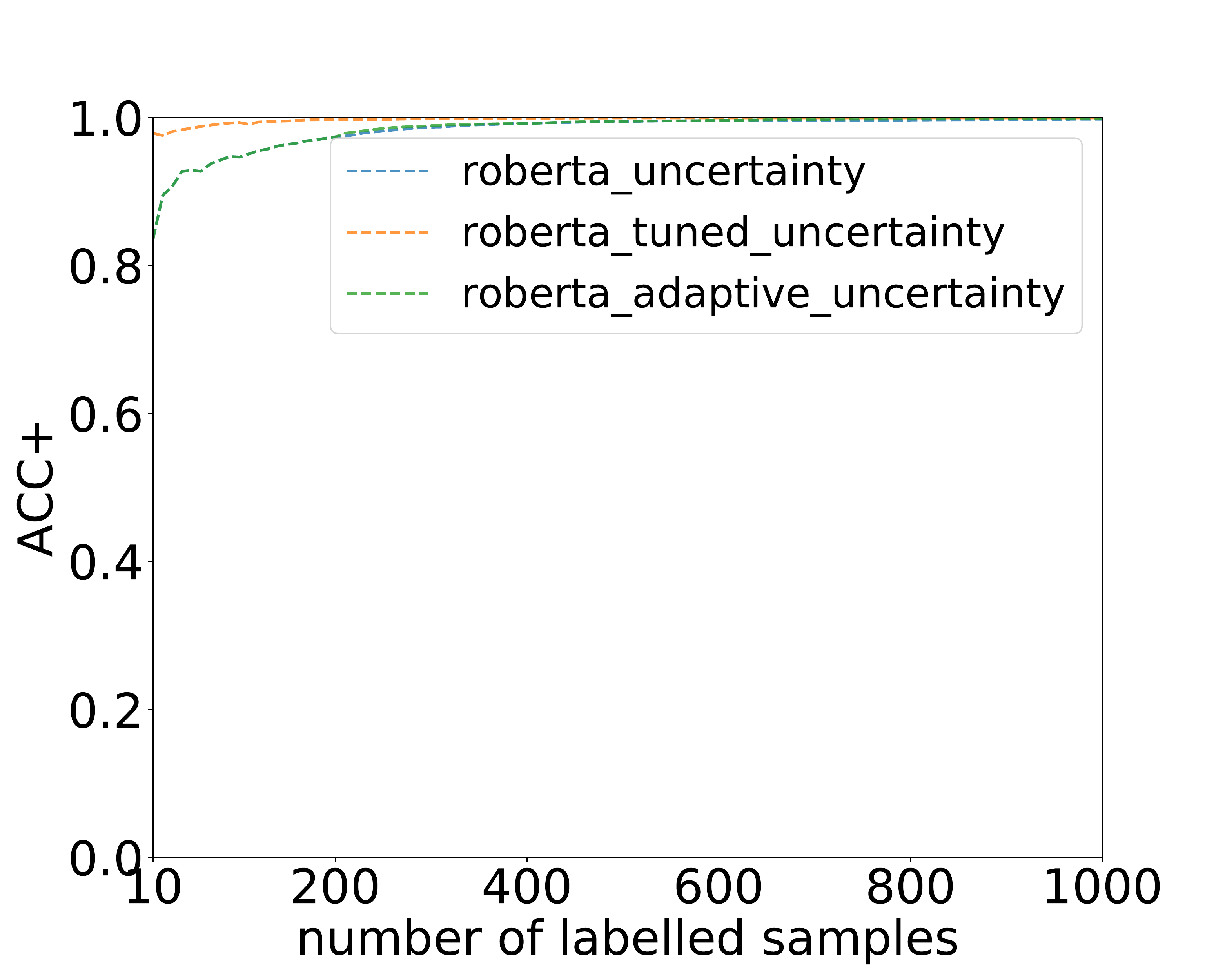}}
\medskip
  \centering
  \subfigure[Additional Customer Review]{\includegraphics[width=0.24\linewidth]{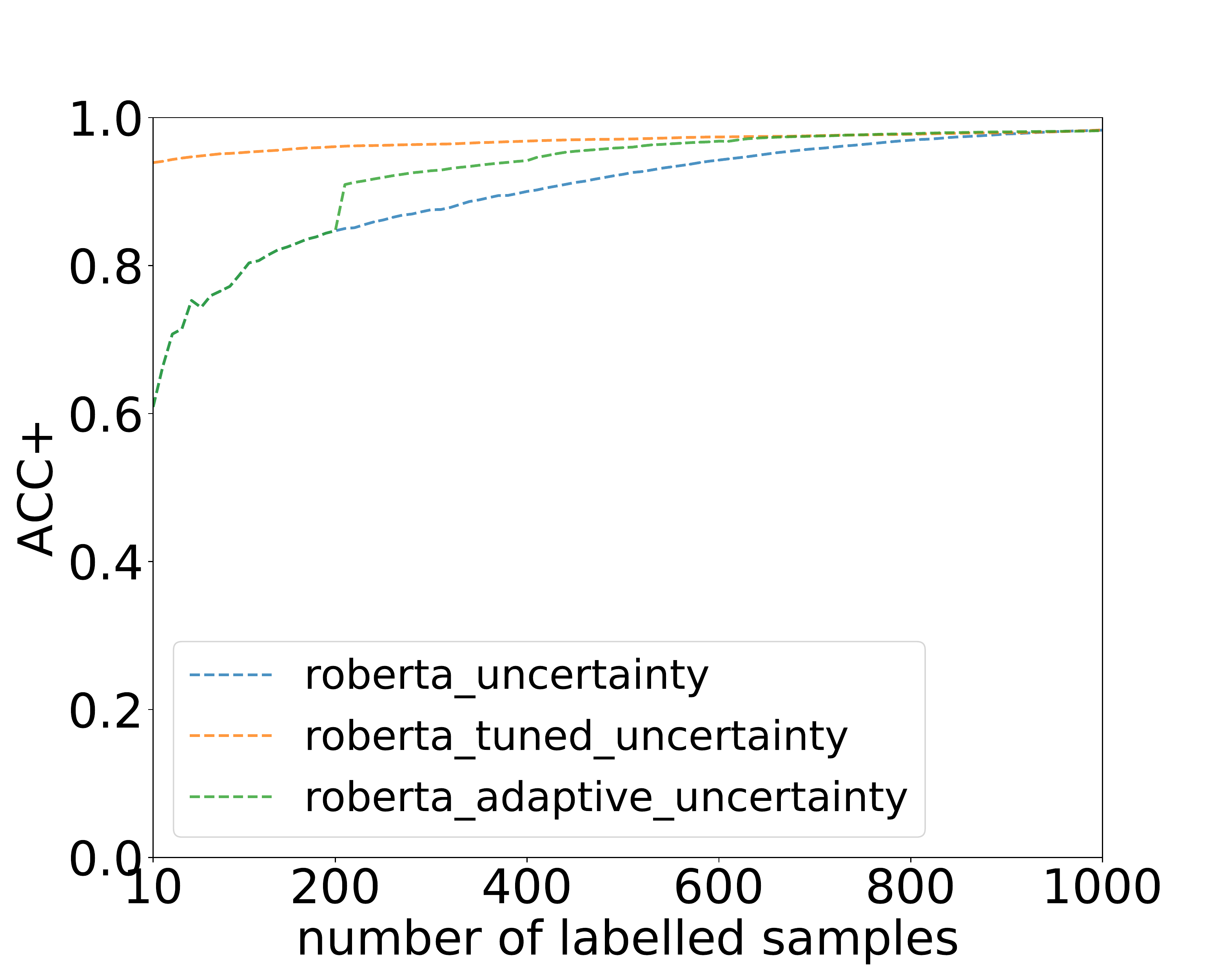}}
  \subfigure[Movie Review Subjectivity]{\includegraphics[width=0.24\linewidth]{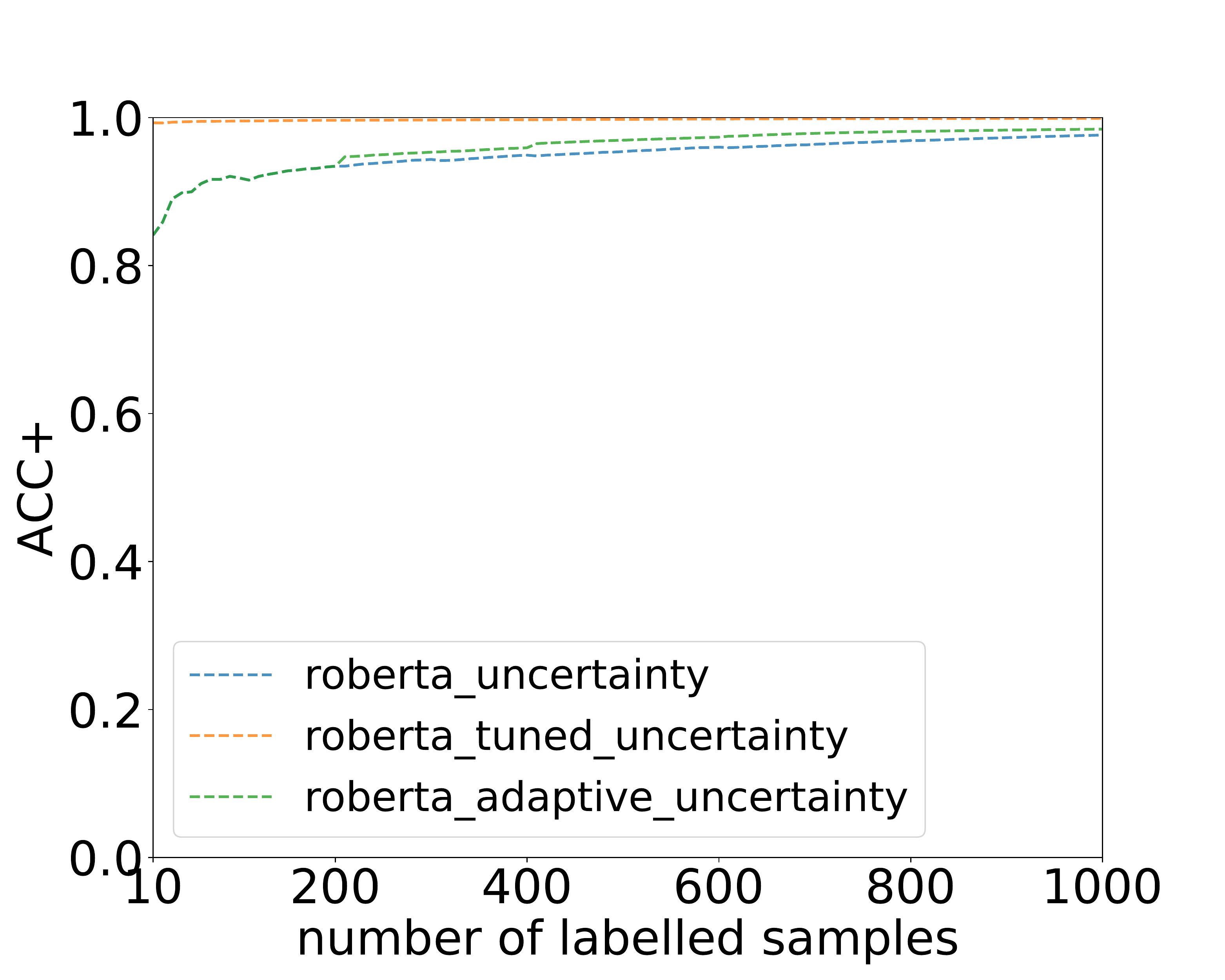}}
  \subfigure[Ag News]{\includegraphics[width=0.24\linewidth]{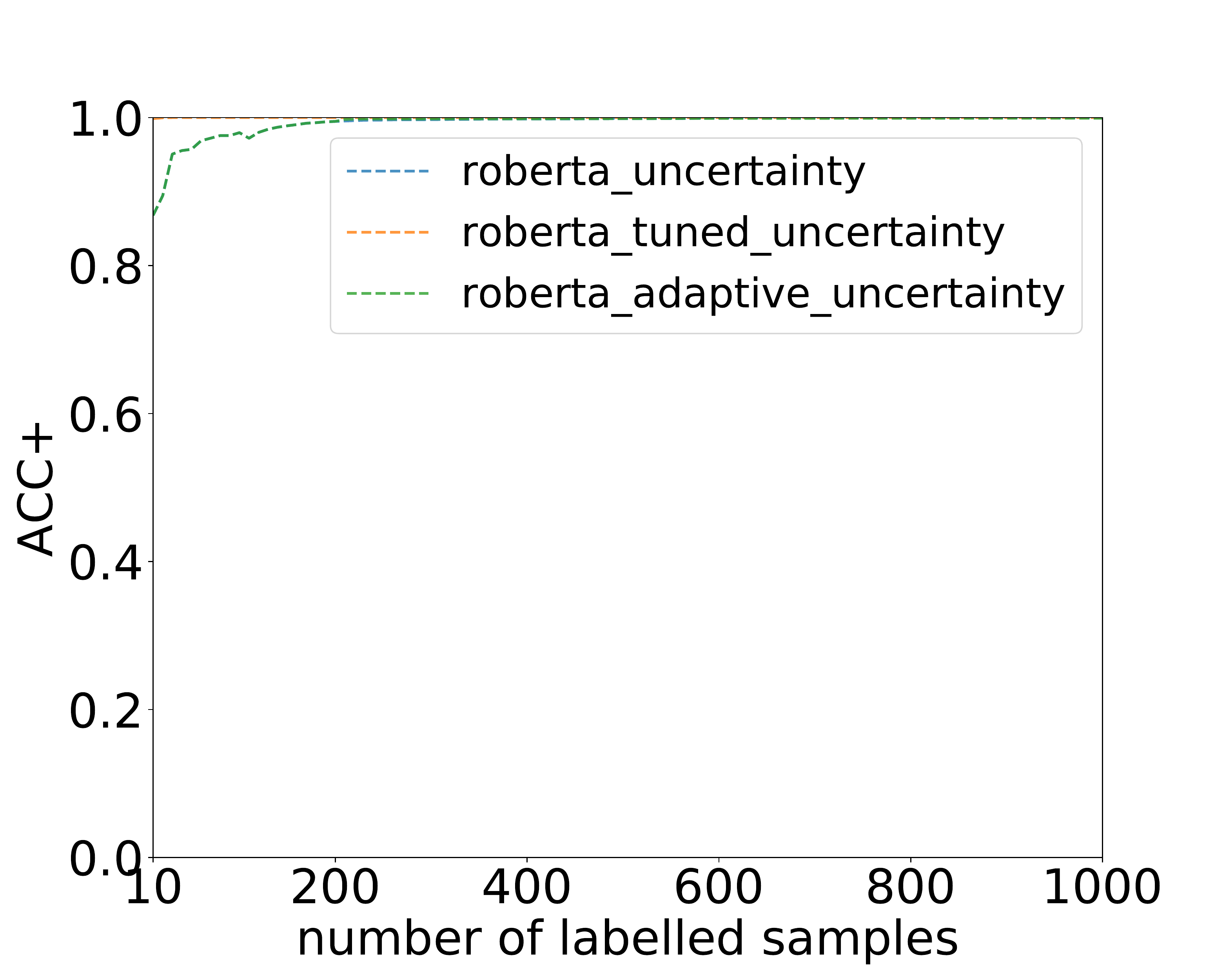}}
  \subfigure[Dbpedia]{\includegraphics[width=0.24\linewidth]{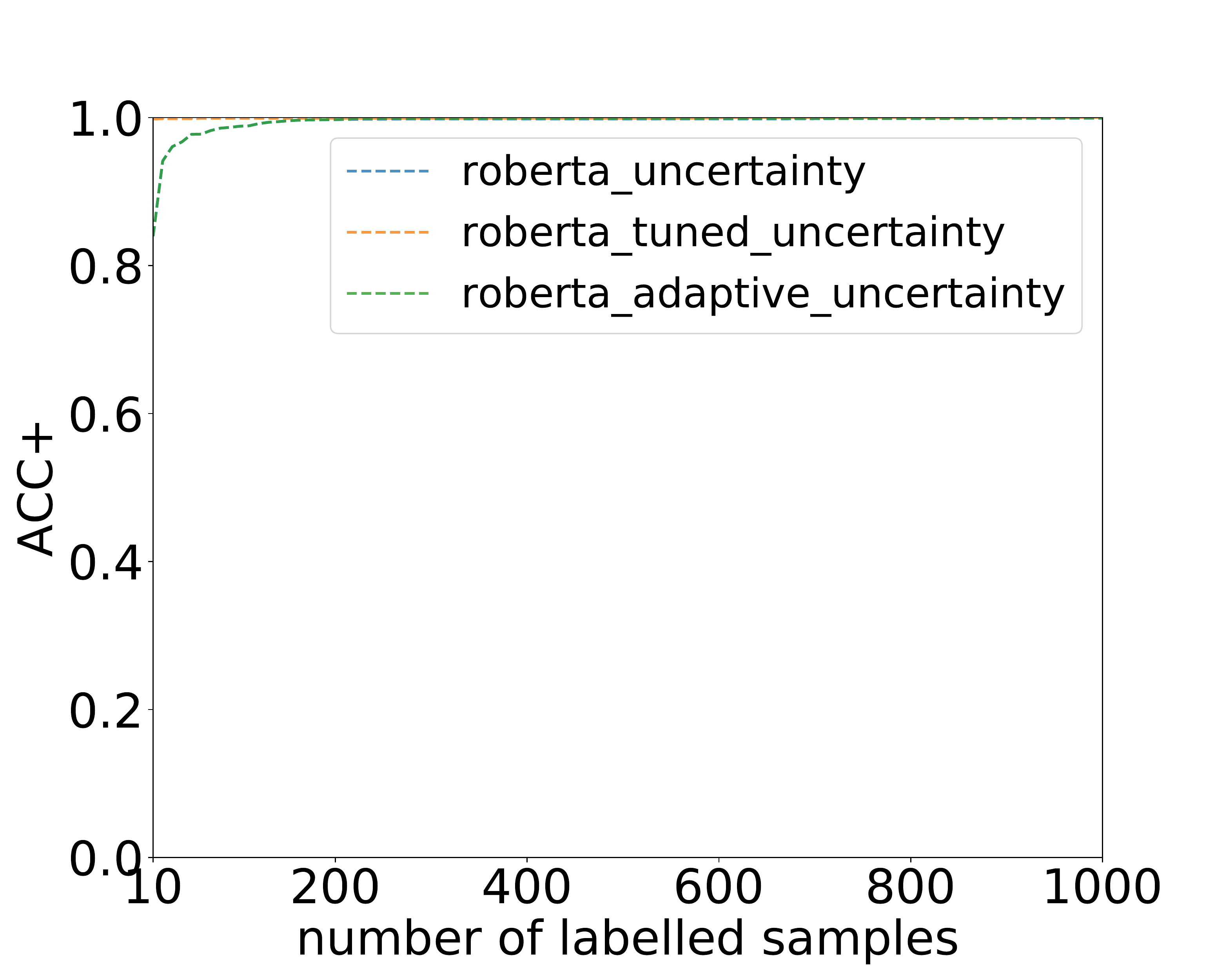}}
  \caption{Comparisons of performance of active learning process with plain pre-trained Roberta, Adaptive Tuning Roberta and fully fine-tuned Roberta. The X-axis represents the number of documents that have been manually annotated and the Y-axis denotes accuracy+. Each curve starts with 10 along the X-axis because of the seeded instances.}
  \label{fig:ATAL}
  
\end{figure*}

Figure \ref{fig:ATAL} illustrate the comparison results of our proposed ATAL algorithm with the traditional active learning process. The orange dotted line denotes the Roberta-model fine-tuned with the fully labelled dataset, which gives us an upper bound on performance for each dataset. The performance of fully fine-tuned Roberta is always close to 1 which is corresponding to our expectation, since we used the whole fully labelled dataset to fine-tune it. The blue and green dotted lines indicate the learning curve of the normal Roberta model and ATAL Roberta model respectively. Notably, in five of eight dataset, the ATAL approach consistently outperforms the more basic approach by a big margin, especially on the Multi-domain Customer Review and Additional Customer Review datasets. The differences appeared when 200 instanced labelled (the step of 20) where our ATAL does its first fine tuning, which proves the effectiveness of our algorithm. In Guardian 2013, AG News and Dbpedia, the performances of the two methods show no difference which is because these three datasets are so easy for the pre-trained Roberta model that the accuracy+ score reaches 1 before ATAL performs any fine tuning. However, this also demonstrates that, at least, the ATAL algorithm does not hurt the performance.

\section{Conclusions}\label{sec:conclusions}

Active learning processes used with text data rely heavily on the document representation mechanism used. This paper presented an evaluation experiment which explored the effectiveness of different text representations in an active learning context. The performance of different text representation techniques combined with popular selection strategies was compared over datasets from different domains. The comparison showed that the transformer-based-embeddings, which are rarely used in active learning, lead to better performance compared to vector based representations such as BOW or simpler word embeddings. Several of the most commonly used selection strategies have been applied in experiments to mitigate the impact of specific selection strategies on the effectiveness of different text representations. Notably, Roberta combined with uncertainty sampling greatly facilitates the application of active learning for text labelling. 

Since several BERT-like models has shown its great performance in active learning a study is carried out to investigate the effectiveness of various BERT-like models and identify the appropriate representation method for pre-trained language models. Our experiments show that representations based on Roberta seem to be best, while DistilBert offers competitive performance with a much lower computational burden. In addition, the averaged word representations is the most effective compared with ``[CLS]'' token representations. 

Lastly, we proposed the Adaptive Tuning Active Learning (ATAL) algorithm where labelled information is fully utilized not only for training classifiers but also for further improving the effectiveness of embeddings produced. We suggest that Robert + averaged representations + uncertainty as the default setting of active learning. If the user has problem with GPU memory, DistilBert or Albert could be another option. Besides, ATAL based on Roberta + uncertainty can be considered due to its great performance shown in the experiment.

While some of the findings in this study may not be that surprising (the fact that representations based on transformer-based models are very effective), this paper makes an important contribution as a comprehensive evaluation experiment showing the effectiveness of different representations.  

An important application of active learning is to labelling the included/excluded studies in literature review \citep{wallace2010active,hashimoto2016topic,miwa2014reducing} which is usually an imbalanced dataset. So it leads to more exploration of the active learning framework over an imbalanced dataset in future work.




\section*{Acknowledgements}
This work was supported by the Teagasc Walsh Scholarship Programme Reference Number [201603]; and the Science Foundation Ireland (SFI) under Grant Number [SFI/12/RC/2289].

\bibliographystyle{modelsarticle-harv}
\bibliography{eswa2020_investigating} 


\end{document}